\documentclass{article}

\usepackage{epsfig}

\newcommand{\ba}{\begin{eqnarray}}
\newcommand{\ea}{\end{eqnarray}}
\def\ii{\'{\i}}
\begin{document}
\pagestyle{plain}

\title{The structure of the nucleon: baryons and pentaquarks
\footnote{Lectures notes, `IV Escuela Mexicana de F\ii sica Nuclear',  
M\'exico, Distrito Federal, June 28 - July 10, 2005}} 
\author{Roelof Bijker\\
Instituto de Ciencias Nucleares,\\ 
Universidad Nacional Aut\'onoma de M\'exico,\\
A.P. 70-543, 04510 M\'exico, Distrito Federal, M\'exico}
\date{April 27, 2005}
\maketitle

\begin{abstract}
In these lecture notes I discuss an algebraic model of baryons 
and pentaquarks in which the permutation symmetry among the 
quarks is taken into account exactly. In particular, a stringlike 
collective model is considered in which the radial excitations of 
baryons and pentaquarks are interpreted as rotations and vibrations 
of the strings. The algebraic structure of the model makes it possible 
to derive closed expressions for physical observables, such as masses 
and electromagnetic and strong couplings. The model is applied to 
the mass spectrum and magnetic moments of baryon resonances of the 
nucleon and delta families and to exotic baryons of the $\Theta$ family. 
The ground state pentaquark is predicted to have angular momentum and 
parity $J^P=1/2^-$ and a small magnetic moment of 0.382 $\mu_N$. 
\end{abstract}

\section{Introduction} 

The structure of the nucleon is of fundamental importance in 
nuclear and particle physics. 
The first indication that the nucleon is not a point particle but has 
an internal structure came from the measurement of the anomalous 
magnetic moment of the proton in the 1930's \cite{stern}, which 
was determined to be 2.5 times as large as one would expect for 
a spin $1/2$ Dirac particle (the actual value is 2.793). The finite 
size of the proton was measured in the 1950's in electron scattering 
experiments at SLAC to be $\sim 0.8$ fm \cite{hofstadter} (compared to the 
current value of 0.895 fm). The first evidence for point-like constituents 
(quarks) inside the proton was found in deep-inelastic-scattering 
experiments in the late 1960's by the MIT-SLAC collaboration \cite{dis} 
which eventually, together with many other developments, would lead to the 
formulation of QCD in the 1970's as the theory of strongly interacting 
particles. 
The complex structure of the proton manifested itself once again in 
recent polarization transfer experiments \cite{jones} which showed 
that the ratio of electric and magnetic form factors of the proton 
exhibits a dramatically different behavior as a function of the momentum 
transfer as compared to the generally accepted picture of form factor 
scaling obtained from the Rosenbluth separation method \cite{dejager}. 

The building blocks of atomic nuclei, the nucleons, are composite 
extended objects. High precision data on the properties of the nucleon 
and its excited states, collectively known as baryons, have been 
accumulated over the past years at Jefferson Laboratory, MIT-Bates, 
LEGS at BNL, MAMI in Mainz, ELSA in Bonn, GRAAL in Grenoble and LEPS 
in Osaka \cite{burkert}. 
To first approximation, the internal structure of the nucleon 
at low energy can be ascribed to three bound constituent quarks $q^3$. 
The baryons are accommodated into flavor singlets, octets and decuplets 
\cite{eightfold}. Each flavor multiplet consists of families of baryons 
characterized by their isospin and strangeness.   
The strangeness of the known baryons is either zero (nucleon, $\Delta$) 
or negative ($\Lambda$, $\Sigma$, $\Xi$ and $\Omega$). Baryons with quantum 
numbers that cannot be obtained from triplets of quarks are called exotic. 

Until recently, there was no experimental evidence for the existence of such 
exotic baryons. The discovery of the $\Theta(1540)$ baryon with positive 
strangeness ${\cal S}=+1$ by the LEPS Collaboration \cite{leps} as the 
first example of an exotic baryon, and the subsequent confirmation by 
various other experimental collaborations has sparked an enormous amount of 
experimental and theoretical studies of exotic baryons \cite{expenta,thpenta}. 
The width of this state is observed to be very small $< 20$ MeV (or 
perhaps as small as a few MeV's). There have also been reports in 
which the pentaquark signal is attributed to kinematical reflections from 
the decay of mesons \cite{dzierba}, or in which no evidence has been found 
for such states \cite{expenta}.  
In addition, evidence has been reported for an exotic baryon 
$\Xi^{--}(1862)$ with strangeness ${\cal S}=-2$ \cite{cern} and for a 
heavy pentaquark $\Theta_c(3099)$ \cite{h1}. The latter two have not been 
confirmed by other experimental collaborations. 

In these lecture notes, I discuss some properties of baryon resonances 
and pentaquarks in a stringlike collective model in which the baryons 
(three-quark or pentaquark) are interpreted as rotations and vibrations 
of the strings. First, in Section 2 some general aspects of multiquark 
states are discussed which are applied to three-quark baryons in Section 3 
and to pentaquarks in Section 4. A summary and conclusions are presented in 
Section 5. Some technical details concerning the spin and flavor wave 
functions are discussed in the appendices.

\section{Multiquark states}

Multiquark states depend both on the internal degrees of freedom 
of color, flavor and spin and the spatial degrees of freedom.   
The classification of the states will be studied from symmetry 
principles without introducing an explicit dynamical model. 
The construction of the classification scheme is guided by two 
conditions: the total multiquark wave function should be a color 
singlet and should be antisymmetric under any permutation of the quarks.  

The internal degrees of freedom are taken to be the three light flavors 
$u$, $d$, $s$ with spin $S=1/2$ and three possible colors $r$, $g$, $b$.   
The internal algebraic structure of the constituent parts consists of the 
usual spin-flavor ($\rm sf$) and color ($\rm c$) algebras
\ba
{\cal G}_{\rm sfc} = SU_{\rm sf}(6) \otimes SU_{\rm c}(3) ~, 
\label{gint}
\ea
where the $SU_{\rm c}(3)$ algebra decribes the (unitary) transformations 
among the three different colors. The spin-flavor algebra can be decomposed 
into 
\ba
SU_{\rm sf}(6) \supset SU_{\rm f}(3) \otimes SU_{\rm s}(2) ~, 
\ea
where the $SU_{\rm f}(3)$ algebra decribes the transformations among 
the three different flavors and $SU_{\rm s}(2)$ among the two 
spin states of the quarks. 
The flavor algebra in turn can be decomposed into 
\ba
SU_{\rm f}(3) \supset SU_{\rm I}(2) \otimes U_{\rm Y}(1) ~, 
\ea
where $I$ denotes the isospin and $Y$ the hypercharge of the quarks. 
The states of a given flavor multiplet  
can be labeled by isospin $I$, $I_3$ and hypercharge $Y$. 
The electric charge is given by the Gell-Mann-Nishijima relation 
\ba
Q \;=\; I_3 + \frac{Y}{2} \;=\; I_3 + \frac{B+{\cal S}}{2} ~, 
\label{GMN}
\ea 
where $B$ denotes the baryon number and ${\cal S}$ the strangeness.   
The quantum numbers of the three light quarks and antiquarks are given in 
Table~\ref{quarks}. The quarks have baryon number $B=1/3$ and spin and 
parity $S^P=1/2^+$ whereas the antiquarks have $B=-1/3$ and 
$S^P=1/2^-$.  

\begin{table}
\centering
\caption[]{\small Quantum number of the quarks and antiquarks} 
\label{quarks}
\vspace{15pt}
\begin{tabular}{crcccrrrr}
\hline
& & & & & & & & \\
& $B$ & $S$ & $P$ & $I$ & $I_3$ & ${\cal S}$ & $Y$ & $Q$ \\
& & & & & & & & \\
\hline
& & & & & & & & \\
$u$ & $ 1/3$ & $1/2$ & $+$ & $1/2$ & $ 1/2$ 
& $ 0$ & $ 1/3$  & $ 2/3$ \\ 
$d$ & $ 1/3$ & $1/2$ & $+$ & $1/2$ & $-1/2$ 
& $ 0$ & $ 1/3$  & $-1/3$ \\ 
$s$ & $ 1/3$ & $1/2$ & $+$ & $0$ & $0$ 
& $-1$ & $-2/3$  & $-1/3$ \\ 
& & & & & & & & \\
$\bar{u}$ & $-1/3$ & $1/2$ & $-$ & $1/2$ & $-1/2$ 
& $0$ & $-1/3$  & $-2/3$ \\ 
$\bar{d}$ & $-1/3$ & $1/2$ & $-$ & $1/2$ & $ 1/2$ 
& $0$ & $-1/3$  & $ 1/3$ \\ 
$\bar{s}$ & $-1/3$ & $1/2$ & $-$ & $0$ & $0$ 
& $1$ & $ 2/3$  & $ 1/3$ \\ 
& & & & & & & & \\
\hline 
\end{tabular}
\end{table}

I shall make use of the Young tableau technique \cite{Hamermesh} to 
construct the allowed representations of $SU(n)$ for the multiquark 
system with $n=2$, $3$ and $6$ for the spin, flavor (or color) and 
spin-flavor degrees of freedom, respectively. 
The fundamental representation of $SU(n)$ is denoted by a box.  
The Young tableaux of $SU(n)$ are labeled by a string of $n$ numbers 
$[f_1,f_2,\ldots,f_n]$ with $f_1 \geq f_2 \geq \ldots \geq f_n$ where   
$f_i$ denotes the number of boxes in the $i$-th row. The labels which are 
zero are usually not written explicitly. The quarks transform as the 
fundamental representation $[1]$ under $SU(n)$, whereas the antiquarks  
transform as the conjugate representation $[1^{n-1}]$ under $SU(n)$ 
\cite{Close,Stancu}. 
As a consequence, the three quarks belong to the flavor 
triplet $[1]$ of $SU_{\rm f}(3)$ and the three antiquarks to the 
antitriplet $[11]$. The spin of the quarks and the antiquarks is determined 
by the representation $[f_1,f_2]$ of $SU_{\rm s}(2)$ as $S=(f_1-f_2)/2$. 
The spin-flavor classification of a single quark and antiquark is given by 
\ba
\begin{array}{cccccc}
& SU_{\rm sf}(6) & \supset & SU_{\rm f}(3) & \otimes & SU_{\rm s}(2) \\
& & & & & \\ 
\mbox{quark} & $[1]$ & \supset & $[1]$ & \otimes & $[1]$ \\
& & & & & \\ 
& \setlength{\unitlength}{1.0pt}
\begin{picture}(10,10)(0,0)
\thinlines
\put ( 0, 0) {\line (1,0){10}}
\put ( 0,10) {\line (1,0){10}}
\put ( 0, 0) {\line (0,1){10}}
\put (10, 0) {\line (0,1){10}}
\end{picture} & \supset & 
\setlength{\unitlength}{1.0pt}
\begin{picture}(10,10)(0,0)
\thinlines
\put ( 0, 0) {\line (1,0){10}}
\put ( 0,10) {\line (1,0){10}}
\put ( 0, 0) {\line (0,1){10}}
\put (10, 0) {\line (0,1){10}}
\end{picture} & \otimes &
\setlength{\unitlength}{1.0pt}
\begin{picture}(10,10)(0,0)
\thinlines
\put ( 0, 0) {\line (1,0){10}}
\put ( 0,10) {\line (1,0){10}}
\put ( 0, 0) {\line (0,1){10}}
\put (10, 0) {\line (0,1){10}}
\end{picture} \\
& & & & & \\
\mbox{antiquark} & $[11111]$ & \supset & $[11]$ & \otimes & $[1]$ \\
& & & & & \\
& \setlength{\unitlength}{1.0pt}
\begin{picture}(10,10)(0,0)
\thinlines
\put ( 0, 10) {\line (1,0){10}}
\put ( 0,  0) {\line (1,0){10}}
\put ( 0,-10) {\line (1,0){10}}
\put ( 0,-20) {\line (1,0){10}}
\put ( 0,-30) {\line (1,0){10}}
\put ( 0,-40) {\line (1,0){10}}
\put ( 0,-40) {\line (0,1){50}}
\put (10,-40) {\line (0,1){50}}
\end{picture} & \supset & 
\setlength{\unitlength}{1.0pt}
\begin{picture}(10,10)(0,0)
\thinlines
\put ( 0, 10) {\line (1,0){10}}
\put ( 0,  0) {\line (1,0){10}}
\put ( 0,-10) {\line (1,0){10}}
\put ( 0,-10) {\line (0,1){20}}
\put (10,-10) {\line (0,1){20}}
\end{picture} & \otimes &
\setlength{\unitlength}{1.0pt}
\begin{picture}(10,10)(0,0)
\thinlines
\put ( 0, 0) {\line (1,0){10}}
\put ( 0,10) {\line (1,0){10}}
\put ( 0, 0) {\line (0,1){10}}
\put (10, 0) {\line (0,1){10}}
\end{picture} \\
& & & & & \\
\end{array}
\label{qqbar}
\ea

\

\

\

\noindent
The spin-flavor states of multiquark systems can be obtained by taking 
the outer product of the representations of the quarks and/or antiquarks 
\cite{jaffe,mulders}. 

The requirement that physical states be color singlets, makes the quarks 
cluster into three-quark triplets ($qqq$ baryons), quark-antiquark 
pairs ($q \bar{q}$ mesons) or products thereof. In general, the 
multiquark configurations can be expressed as 
\ba
q^{3m+n} \bar{q}^n ~,
\ea
which reduces to $qqq$ baryons for $m=1$ and $n=0$ and to $q \bar{q}$ 
mesons for $m=0$ and $n=1$. In these lecture notes, I consider $q^3$ 
baryons and $q^4 \bar{q}$ pentaquarks ($m=n=1$). The latter were first 
considered in the late 70's \cite{RLJ,sorba,DS}. 

\section{$q^3$ Baryons}

The nucleon is not an elementary particle, but it is generally viewed as a 
confined system of three constituent quarks interacting via gluon exchange. 
Effective models of the nucleon and its excited states (or baryon resonances) 
are based on three constituent parts that carry the internal degrees 
of freedom of spin, flavor and color, but differ in their treatment of radial 
(or orbital) excitations. 

In this section I discuss the well-known example of $qqq$ baryons. 
Baryons are considered to be built of three constituent quarks which are 
characterized by both internal and spatial degrees of freedom. 

\subsection{Internal degrees of freedom}

\begin{table}[b]
\centering
\caption[]{\small Allowed color, spin, flavor and spin-flavor baryon states}
\vspace{15pt}
\label{bstates}
$\begin{array}{cccc}
\hline
& & & \\
& q^3 & \mbox{Dimension} & S_3 \sim D_3 \\
& & & \\
\hline
& & & \\
\mbox{color} & [111] & \mbox{singlet} & A_2 \\
& & & \\
\hline
& & & \\
\mbox{spin} & [3] & 4 & A_1 \\
& [21] & 2 & E \\
& & & \\
\hline
& & & \\
\mbox{flavor} & [3] & \mbox{decuplet} & A_1 \\
& [21] & \mbox{octet} & E \\
& [111] & \mbox{singlet} & A_2 \\
& & & \\
\hline
& & & \\
\mbox{spin-flavor} 
& [3]   & 56 & A_1 \\
& [21]  & 70 & E   \\
& [111] & 20 & A_2 \\
& & & \\
\hline
\end{array}$
\end{table}

The allowed spin, flavor and spin-flavor states are obtained by 
standard group theoretic techniques \cite{Hamermesh,Close,Stancu}. 
For example, the total spin of the three-quark system is obtained by coupling 
the three $1/2$ spins to give $S=3/2$ and $S=1/2$ (twice). In general, 
the spin, flavor and spin-flavor states of the three-quark system 
are obtained by taking the product 
\ba
\begin{array}{ccccccccccc}
[1] &\otimes& [1] &\otimes& [1] &=& [3] &\oplus& 2 \; [21] &\oplus& [111] \\ 
\setlength{\unitlength}{1.0pt}
\begin{picture}(10,10)(0,0)
\thinlines
\put ( 0, 0) {\line (1,0){10}}
\put ( 0,10) {\line (1,0){10}}
\put ( 0, 0) {\line (0,1){10}}
\put (10, 0) {\line (0,1){10}}
\end{picture} &\otimes& 
\setlength{\unitlength}{1.0pt}
\begin{picture}(10,10)(0,0)
\thinlines
\put ( 0, 0) {\line (1,0){10}}
\put ( 0,10) {\line (1,0){10}}
\put ( 0, 0) {\line (0,1){10}}
\put (10, 0) {\line (0,1){10}}
\end{picture} &\otimes& 
\setlength{\unitlength}{1.0pt}
\begin{picture}(10,10)(0,0)
\thinlines
\put ( 0, 0) {\line (1,0){10}}
\put ( 0,10) {\line (1,0){10}}
\put ( 0, 0) {\line (0,1){10}}
\put (10, 0) {\line (0,1){10}}
\end{picture} &=& 
\setlength{\unitlength}{1.0pt}
\begin{picture}(30,10)(0,0)
\thinlines
\put ( 0, 0) {\line (1,0){30}}
\put ( 0,10) {\line (1,0){30}}
\put ( 0, 0) {\line (0,1){10}}
\put (10, 0) {\line (0,1){10}}
\put (20, 0) {\line (0,1){10}}
\put (30, 0) {\line (0,1){10}}
\end{picture} &\oplus& 2 \;\;  
\setlength{\unitlength}{1.0pt}
\begin{picture}(20,20)(0,5)
\thinlines
\put ( 0, 0) {\line (1,0){10}}
\put ( 0,10) {\line (1,0){20}}
\put ( 0,20) {\line (1,0){20}}
\put ( 0, 0) {\line (0,1){20}}
\put (10, 0) {\line (0,1){20}}
\put (20,10) {\line (0,1){10}}
\end{picture} &\oplus& 
\setlength{\unitlength}{1.0pt}
\begin{picture}(10,30)(0,10)
\thinlines
\put ( 0, 0) {\line (1,0){10}}
\put ( 0,10) {\line (1,0){10}}
\put ( 0,20) {\line (1,0){10}}
\put ( 0,30) {\line (1,0){10}}
\put ( 0, 0) {\line (0,1){30}}
\put (10, 0) {\line (0,1){30}}
\end{picture} 
\end{array}
\label{qqq}
\ea

\

\noindent
where each of the boxes on the left-hand side denotes a quark. 

\begin{table}
\centering
\caption[]{\small Classification of ground state baryons according 
to $SU_{\rm f}(3) \supset SU_{\rm I}(2) \otimes U_{\rm Y}(1)$} 
\label{baryons} 
\vspace{15pt} 
\begin{tabular}{lllcrc}
\hline
& & & & & \\
& & & $I$ & $Y$ & $Q$ \\
& & & & & \\
\hline
& & & & & \\
$J^P=\frac{1}{2}^+$ octet 
& Nucleon & $N$       & $\frac{1}{2}$ &   1 &     0,1 \\
& Sigma   & $\Sigma$  &  1  &   0 & --1,0,1 \\
& Lambda  & $\Lambda$ &  0  &   0 &     0 \\
& Xi      & $\Xi$     & $\frac{1}{2}$ & --1 & --1,0 \\
& & & & & \\
\hline
& & & & & \\
$J^P=\frac{3}{2}^+$ decuplet 
& Delta & $\Delta$        & $\frac{3}{2}$ &   1 & --1,0,1,2 \\
& Sigma & $\Sigma^{\ast}$ &  1  &   0 & --1,0,1 \\
& Xi    & $\Xi^{\ast}$    & $\frac{1}{2}$ & --1 & --1,0 \\
& Omega & $\Omega$        &  0  & --2 & --1 \\
& & & & & \\
\hline
\end{tabular}
\end{table}

An important ingredient in the construction of baryon wave 
function is the permutation symmetry between the quarks. If some of 
the constituent parts are identical one must construct states and 
operators that transform according to the representations of the 
permutation group (either $S_3$ for three identical parts or $S_2$ 
for two identical parts). Here I discuss states that have 
good permutation symmetry among the three quarks $S_3$. These 
states form a complete basis which can be used for any calculation 
of baryon properties.
The permutation symmetry of the three-quark system is characterized by 
the $S_3$ Young tableaux $[3]$ (symmetric), $[21]$ (mixed symmetric) 
and $[111]$ (antisymmetric) or, equivalently, by the irreducible 
representations of the point group $D_3$ (which is isomorphic to $S_3$) 
as $A_1$, $E$ and $A_2$, respectively. For notational purposes the 
latter is used to label the discrete symmetry of the baryon wave functions. 
The corresponding dimensions are 1, 2 and 1. 

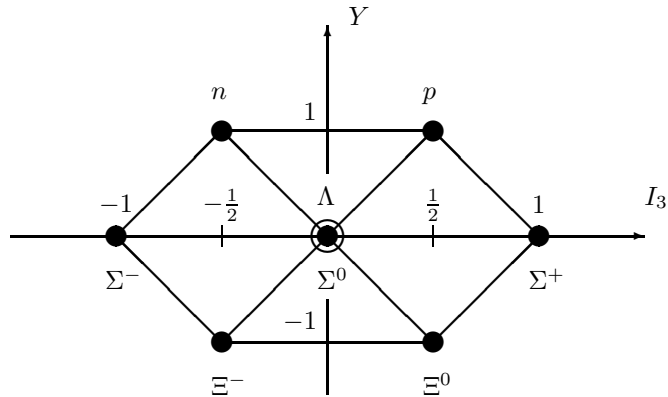
\begin{figure}[b]
\centering
\setlength{\unitlength}{0.8pt}
\begin{picture}(350,200)(75,75)
\thinlines
\put(100,150) {\vector(1,0){300}}
\put(250, 75) {\line(0,1){45}}
\put(250,180) {\vector(0,1){70}}

\put(150,145) {\line(0,1){10}}
\put(200,145) {\line(0,1){10}}
\put(250,145) {\line(0,1){10}}
\put(300,145) {\line(0,1){10}}
\put(350,145) {\line(0,1){10}}
\thicklines
\put(200,200) {\line(1,0){100}}
\put(150,150) {\line(1,0){200}}
\put(200,100) {\line(1,0){100}}

\put(150,150) {\line(1,1){ 50}}
\put(200,100) {\line(1,1){100}}
\put(300,100) {\line(1,1){ 50}}

\put(200,100) {\line(-1,1){ 50}}
\put(300,100) {\line(-1,1){100}}
\put(350,150) {\line(-1,1){ 50}}

\multiput(200,200)(100,0){2}{\circle*{10}}
\multiput(150,150)(100,0){3}{\circle*{10}}
\multiput(200,100)(100,0){2}{\circle*{10}}

\put(250,150){\circle{15}}

\put(195,215){$n$}
\put(295,215){$p$}
\put(145,125){$\Sigma^-$}
\put(245,125){$\Sigma^0$}
\put(245,165){$\Lambda$}
\put(345,125){$\Sigma^+$}
\put(195, 75){$\Xi^-$}
\put(295, 75){$\Xi^0$}

\put(400,165){$I_3$}
\put(260,250){$Y$}
\put(150,165){\makebox(0,0){$-1$}}
\put(200,165){\makebox(0,0){$-\frac{1}{2}$}}
\put(300,165){\makebox(0,0){$ \frac{1}{2}$}}
\put(350,165){\makebox(0,0){$ 1$}}
\put(245,205){\makebox(0,0)[br]{$1$}}
\put(245,105){\makebox(0,0)[br]{$-1$}}
\end{picture}
\caption[]{\small Baryon octet}
\label{octet}
\end{figure}

For the coupling of the spins, the antisymmetric representation $[111]$ 
in Eq.~(\ref{qqq}) does not occur, since the representations of 
$SU_{\rm s}(2)$ can have at most two rows. This means that the spin 
of the three-quark system can be either $S=(f_1-f_2)/2=3/2$ (Young 
tableau $[3]$) or $S=1/2$ (twice, Young tableau $[21]$) with  
permutation symmetry $A_1$ and $E$, respectively. The dimension of 
the spin $S$ is given by the number of spin projections $2S+1$. 
The allowed spin, flavor and spin-flavor states are summarized 
in Table~\ref{bstates}. 
The allowed flavor states are $[3]$, $[21]$ and $[111]$ which are 
usually denoted by their dimensions as 10 (decuplet), 8 (octet) 
and 1 (singlet), respectively. The corresponding point group symmetries 
are $A_1$, $E$ and $A_2$. Finally, the spin-flavor states are  
denoted by their dimensions as $[56]$, $[70]$ and $[20]$ with 
symmetries $A_1$, $E$ and $A_2$, respectively. 

The spin and flavor content of each spin-flavor multiplet is given 
by the decomposition of the representations of $SU_{\rm sf}(6)$ into 
those of $SU_{\rm f}(3) \otimes SU_{\rm s}(2)$ 
\ba
\, [56] &\supset& ^{2}8 \;\oplus\; ^{4}10 ~,
\nonumber\\
\, [70] &\supset&�^{2}8 \;\oplus\; ^{4}8 \;\oplus\; ^{2}10  
\;\oplus\; ^{2}1 ~,
\nonumber\\
\, [20] &\supset&�^{2}8 \;\oplus\; ^{4}1 ~,
\label{qqq2}
\ea
where the superscript denotes $2S+1$. For example, the symmetric 
representation $[56]$ contains an octet with $S=1/2$ and a decuplet 
with $S=3/2$.
 
In Table~\ref{baryons} I present the classification of the baryon 
flavor octet and decuplet in terms of the isospin $I$ and the 
hypercharge $Y$ according to the decomposition of the flavor symmetry 
$SU_{\rm f}(3)$ into $SU_{\rm I}(2) \otimes U_{\rm Y}(1)$. 
The nucleon and $\Delta$ are nonstrange baryons with ${\cal S}=0$, whereas the 
$\Sigma$, $\Lambda$, $\Xi$ and $\Omega$ hyperons carry 
strangeness ${\cal S}=-1$, $-1$, $-2$ and $-3$, respectively. 
The flavor singlet $[111]$ with $A_2$ symmetry consists of a single 
baryon ($\Lambda^{\ast}$) which has isospin $I=0$ and hypercharge $Y=0$ 
(strangeness ${\cal S}=-1$). 

\begin{figure}
\centering
\setlength{\unitlength}{0.8pt}
\begin{picture}(400,250)(75,75)
\thinlines
\put( 75,200) {\vector(1,0){375}}
\put(250,100) {\line(0,1){70}}
\put(250,190) {\vector(0,1){110}}
\put(100,195) {\line(0,1){10}}
\put(150,195) {\line(0,1){10}}
\put(200,195) {\line(0,1){10}}
\put(250,195) {\line(0,1){10}}
\put(300,195) {\line(0,1){10}}
\put(350,195) {\line(0,1){10}}
\put(400,195) {\line(0,1){10}}
\thicklines
\put(100,250) {\line(1,0){300}}
\put(150,200) {\line(1,0){200}}
\put(200,150) {\line(1,0){100}}

\put(150,200) {\line(1,1){ 50}}
\put(200,150) {\line(1,1){100}}
\put(250,100) {\line(1,1){150}}

\put(250,100) {\line(-1,1){150}}
\put(300,150) {\line(-1,1){100}}
\put(350,200) {\line(-1,1){ 50}}

\multiput(100,250)(100,0){4}{\circle*{10}}
\multiput(150,200)(100,0){3}{\circle*{10}}
\multiput(200,150)(100,0){2}{\circle*{10}}
\put(250,100){\circle*{10}}

\put( 95,265){$\Delta^-$}
\put(195,265){$\Delta^0$}
\put(295,265){$\Delta^+$}
\put(395,265){$\Delta^{++}$}
\put(140,175){$\Sigma^{\ast \, -}$}
\put(240,175){$\Sigma^{\ast \, 0}$}
\put(340,175){$\Sigma^{\ast \, +}$}
\put(190,125){$\Xi^{\ast \, -}$}
\put(295,125){$\Xi^{\ast \, 0}$}
\put(245, 75){$\Omega^-$}

\put(450,215){$I_3$}
\put(260,300){$Y$}
\put(100,215){\makebox(0,0){$-\frac{3}{2}$}}
\put(150,215){\makebox(0,0){$-1$}}
\put(200,215){\makebox(0,0){$-\frac{1}{2}$}}
\put(300,215){\makebox(0,0){$ \frac{1}{2}$}}
\put(350,215){\makebox(0,0){$ 1$}}
\put(400,215){\makebox(0,0){$ \frac{3}{2}$}}
\put(245,255){\makebox(0,0)[br]{$1$}}
\put(245,155){\makebox(0,0)[br]{$-1$}}
\end{picture}
\caption[]{\small Baryon decuplet}
\label{decuplet}
\end{figure}
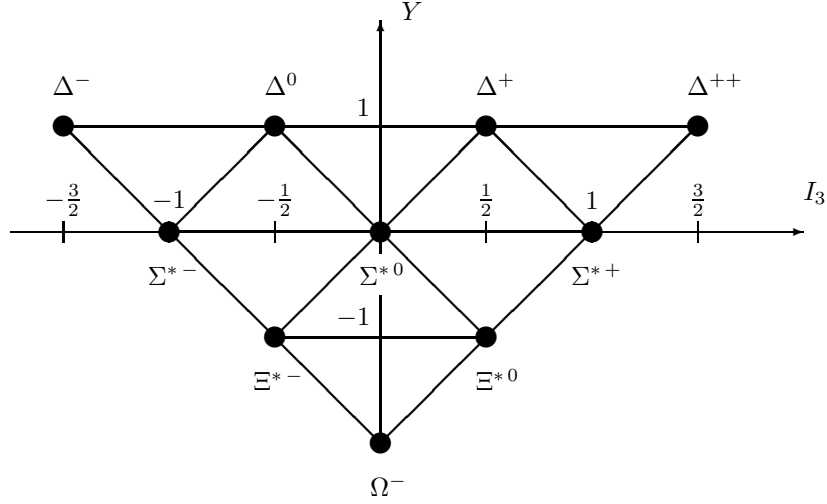

A standard representation of the octet and decuplet baryons is that 
of a socalled weight diagram in the $I_3$-$Y$ plane (see Figs.~\ref{octet} 
and~\ref{decuplet}).

\subsection{Spatial degrees of freedom}

\begin{figure}
\centerline{\hbox{
\epsfig{figure=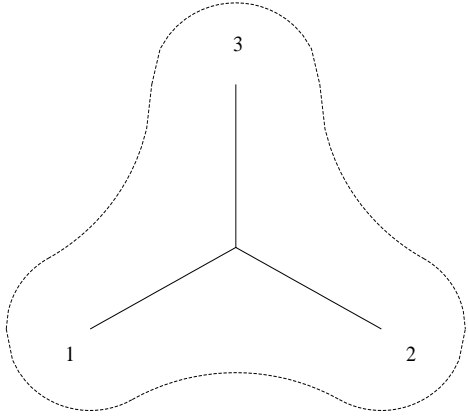,height=0.375\textwidth,width=0.5\textwidth} }}
\caption[]{\small Collective model of baryons}
\label{geometry}
\end{figure}

The relative motion of the three constituent parts can be described in
terms of Jacobi coordinates, $\vec{\rho}$ and $\vec{\lambda}$,
which in the case of three identical objects are 
\ba
\vec{\rho} &=& \frac{1}{\sqrt{2}} (\vec{r}_1 - \vec{r}_2) ~,
\nonumber\\
\vec{\lambda} &=& \frac{1}{\sqrt{6}} (\vec{r}_1 + \vec{r}_2 -2\vec{r}_3) ~, 
\label{jacobi1}
\ea
where $\vec{r}_1$, $\vec{r}_2$ and $\vec{r}_3$ denote the end points of 
the string configuration in Figure~\ref{geometry}. The method of bosonic 
quantization \cite{BIL} consists in introducing a dipole 
boson $b_i^{\dagger}$ with $L^P=1^-$ for each independent relative 
coordinate and its conjugate momentum, and adding an auxiliary scalar 
boson $s^{\dagger}$ with $L^P=0^+$ 
\ba
s^{\dagger} ~, \; b^{\dagger}_{\rho,m} ~, \; b^{\dagger}_{\lambda,m} ~, 
\;\;\; (m=0,\pm 1) ~.   
\label{bb}
\ea
The scalar boson does not represent an independent degree of freedom, 
but is added under the restriction that the total number of bosons 
\ba
\hat N \;=\; s^{\dagger} s + \sum_m \left( b^{\dagger}_{\rho,m} b_{\rho,m} 
+ b^{\dagger}_{\lambda,m} b_{\lambda,m} \right) ~, 
\label{number}
\ea
is conserved. This procedure leads to a compact spectrum generating 
algebra for the radial (or orbital) excitations 
\ba 
{\cal G}_{\rm orb} \;=\; U(7) ~,  \label{u7}
\ea 
which describes the transformations among the seven bosons of Eq.~(\ref{bb}). 
For a system of interacting bosons the model space is spanned by the 
symmetric irreducible representation $[N]$ of $U(7)$. 
The value of $N$ determines the size of the model space. 

The $S_3$ permutation symmetry poses an additional constraint on the 
allowed interaction terms. The scalar boson, $s^{\dagger}$, transforms 
as the symmetric representation, $A_1$, while the two vector bosons, 
$b^{\dagger}_{\rho}$ and $b^{\dagger}_{\lambda}$, transform as the two 
components $E_{\rho}$ and $E_{\lambda}$ of the mixed symmetry representation $E$.  
The choice of the Jacobi coordinates in Eq.~(\ref{jacobi1}) is consistent 
with the conventions used for the spin and flavor wave functions used in 
the appendices. The eigenvalues and corresponding eigenvectors can be 
obtained exactly by diagonalization in an appropriate basis. The radial 
wave functions have, by construction, good angular momentum $L$, 
parity $P$, and permutation symmetry $t=A_1$, $E$, $A_2$. Moreover, 
the total number of bosons $N$ is conserved.  

The mass operator depends both on the spatial and the internal degrees 
of freedom. Whereas in nonrelativistic problems the spectrum is 
obtained by expanding the Hamiltonian in terms of the generators of the 
algebra ${\cal G}_{\rm orb}$, in algebraic models of hadrons one uses the  
mass-squared operator \cite{BIL}. In this section I discuss 
the contribution from the spatial part. The most general form of the 
radial part of the 
mass operator, that preserves angular momentum, parity and the total 
number of bosons, transforms as a scalar under the permutation group 
and is at most two-body in the boson operators, can be written as 
\ba
\hat M^2_{\rm orb} &=& \epsilon_s \, s^{\dagger} \tilde{s} 
- \epsilon_p \, (b_{\rho}^{\dagger} \cdot \tilde{b}_{\rho}
+ b_{\lambda}^{\dagger} \cdot \tilde{b}_{\lambda})
+ u_0 \, (s^{\dagger} s^{\dagger} \tilde{s} \tilde{s}) - u_1 \,
  s^{\dagger} ( b^{\dagger}_{\rho} \cdot \tilde{b}_{\rho}
+ b^{\dagger}_{\lambda} \cdot \tilde{b}_{\lambda} ) \tilde{s}
\nonumber\\
&& + v_0 \, \left[ ( b^{\dagger}_{\rho} \cdot b^{\dagger}_{\rho}
+ b^{\dagger}_{\lambda} \cdot b^{\dagger}_{\lambda} ) \tilde{s} \tilde{s} 
+ s^{\dagger} s^{\dagger} ( \tilde{b}_{\rho} \cdot \tilde{b}_{\rho}
+ \tilde{b}_{\lambda} \cdot \tilde{b}_{\lambda} ) \right]
\nonumber\\
&& + \sum_{l=0,2} c_l \, \left[ 
( b^{\dagger}_{\rho} \times   b^{\dagger}_{\rho} 
- b^{\dagger}_{\lambda} \times b^{\dagger}_{\lambda} )^{(l)} \cdot
( \tilde{b}_{\rho} \times \tilde{b}_{\rho}
- \tilde{b}_{\lambda} \times \tilde{b}_{\lambda} )^{(l)}
+ 4 \, ( b^{\dagger}_{\rho} \times b^{\dagger}_{\lambda})^{(l)} \cdot
       ( \tilde b_{\lambda} \times \tilde b_{\rho})^{(l)} \right]
\nonumber\\
&& + c_1 \, ( b^{\dagger}_{\rho} \times b^{\dagger}_{\lambda} )^{(1)} 
\cdot ( \tilde b_{\lambda} \times \tilde b_{\rho} )^{(1)}
+ \sum_{l=0,2} w_l \, ( b^{\dagger}_{\rho} \times b^{\dagger}_{\rho}
  + b^{\dagger}_{\lambda} \times b^{\dagger}_{\lambda} )^{(l)} \cdot
  ( \tilde{b}_{\rho} \times \tilde{b}_{\rho}
  + \tilde{b}_{\lambda} \times \tilde{b}_{\lambda} )^{(l)} ~,
\nonumber\\ \label{ms3}
\ea
with $\tilde{s}=s$ and $\tilde{b}_{i,m}=(-1)^{1-m}b_{i,-m}$. 
Here the dots indicate scalar products and the crosses tensor products 
with respect to the rotation group. 
The eigenvalues and corresponding eigenvectors of the mass-squared 
operator of Eq.~(\ref{ms3}) can be obtained exactly by numerical 
diagonalization. The wave functions obtained in this way have by construction 
good angular momentum, parity and permutation symmetry. The procedure to 
determine the permutation symmetry of a given wave function is described in  
\cite{BIL}. 

The mass-squared operator of Eq.~(\ref{ms3}) contains several models of 
baryon structure which arise for special choices of the coefficients. 
In the next sections two special solutions are discussed: the harmonic 
oscillator quark model and a stringlike collective model. 

\subsubsection{Harmonic oscillator quark model}

Harmonic oscillator quark models correspond to the choice $v_0=0$, 
{\it i.e.} no coupling between different harmonic oscillator shells. 
The one-body terms of the $S_3$ invariant mass operator of Eq.~(\ref{ms3}) 
correspond to a harmonic oscillator 
\ba
\hat M^2_{\rm orb} \;=\; \epsilon \, \sum_m \left( 
  b^{\dagger}_{\rho,m} b_{\rho,m} 
+ b^{\dagger}_{\lambda,m} b_{\lambda,m} \right) ~, 
\ea
whereas the two-body interactions give rise to anharmonic contributions.   
The nonrelativistic harmonic oscillator quark 
model \cite{IK} is a model of this type, although it is 
written for the mass $\hat M$ rather than for $\hat M^2$. 
The equality of the frequencies of the $\rho$ and $\lambda$ oscillators 
is a consequence of the $S_3$ permutation symmetry. 
The mass spectrum is that of a six-dimensional harmonic oscillator
\ba
M^2_{\rm orb} \;=\; \epsilon \, \left( n_{\rho}+n_{\lambda} \right) ~,
\label{ho1}
\ea
where $n_{\rho}+n_{\lambda}=n$ is the number of oscillator quanta. 
The model space consists of the oscillator shells with $n=0,1,\ldots,N$. 

\begin{figure}
\centering
\setlength{\unitlength}{1.0pt}
\begin{picture}(310,200)(0,0)
\thinlines
\put (  0,  0) {\line(1,0){310}}
\put (  0,200) {\line(1,0){310}}
\put (  0,  0) {\line(0,1){200}}
\put (310,  0) {\line(0,1){200}}
\thicklines
\put ( 70, 40) {\line(1,0){20}}
\put ( 70,100) {\line(1,0){20}}
\put ( 70,160) {\line(1,0){20}}
\put (110,160) {\line(1,0){20}}
\put (150,160) {\line(1,0){20}}
\put (190,160) {\line(1,0){20}}
\put (230,160) {\line(1,0){20}}
\put ( 80, 40) {\vector(0, 1){60}}
\put ( 80,100) {\vector(0,-1){60}}
\thinlines
\put ( 70, 70) {$\epsilon$}
\put ( 30, 37) {$n=0$}
\put ( 30, 97) {$n=1$}
\put ( 30,157) {$n=2$}
\put ( 92, 37) {$0^+_{A_1}$}
\put ( 92, 97) {$1^-_E$}
\put ( 92,157) {$2^+_{A_1}$}
\put (132,157) {$2^+_E$}
\put (172,157) {$1^+_{A_2}$}
\put (212,157) {$0^+_{A_1}$}
\put (252,157) {$0^+_E$}
\end{picture}
\caption[]{\small 
Schematic representation of the radial excitations of $q^3$ baryons 
in a harmonic oscillator model. The number of bosons is $N=2$.}
\label{hosc1}
\end{figure}
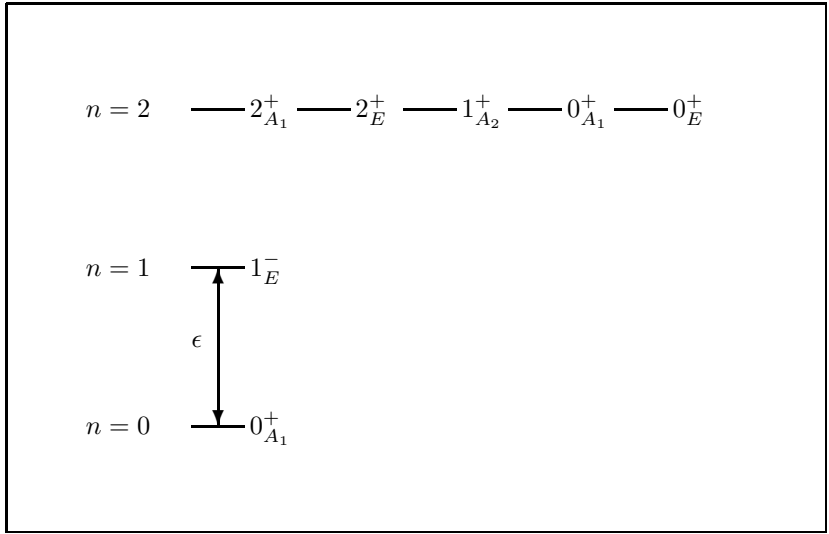

The mass spectrum for the harmonic oscillator is shown in Fig.~\ref{hosc1} 
for $N=2$ bosons. The levels are grouped into oscillator shells characterized 
by $n$. The ground state has $n=0$ and $L^P_t=0^+_{A_1}$. The one-phonon 
multiplet $n=1$ has two degenerate states with $L^P=1^-$ which 
belong to the two-dimensional representation $E$,   
and the two-phonon multiplet $n=2$ consists of the states 
$L^P_t=2^+_{A_1}$, $2^+_{E}$, $1^+_{A_2}$, $0^+_{A_1}$ and $0^+_{E}$. 
The degenerate levels in an oscillator shell can be separated by 
introducing higher-order interactions in the mass operator. 

\subsubsection{Stringlike collective model}

In the stringlike collective model the baryons are interpreted 
as rotational and vibrational excitations of the string configuration 
of Fig.~\ref{geometry}. The three constituent parts move in a correlated way. 
For three identical constituents the vibrations are described by \cite{BIL}
\ba
\hat M^{2}_{\rm vib} &=& \xi_1 \,
\Bigl ( R^2 \, s^{\dagger} s^{\dagger}
- b^{\dagger}_{\rho} \cdot b^{\dagger}_{\rho}
- b^{\dagger}_{\lambda} \cdot b^{\dagger}_{\lambda} \Bigr ) \,
\Bigl ( R^2 \, \tilde{s} \tilde{s} - \tilde{b}_{\rho} \cdot \tilde{b}_{\rho}
- \tilde{b}_{\lambda} \cdot \tilde{b}_{\lambda} \Bigr )
\nonumber\\
&& + \xi_2 \, \Bigl [
\Bigl( b^{\dagger}_{\rho} \cdot b^{\dagger}_{\rho}
- b^{\dagger}_{\lambda} \cdot b^{\dagger}_{\lambda} \Bigr ) \,
\Bigl ( \tilde{b}_{\rho} \cdot \tilde{b}_{\rho}
- \tilde{b}_{\lambda} \cdot \tilde{b}_{\lambda} \Bigr )
+ 4 \, \Bigl ( b^{\dagger}_{\rho} \cdot b^{\dagger}_{\lambda} \Bigr ) \,
\Bigl ( \tilde{b}_{\lambda} \cdot \tilde{b}_{\rho} \Bigr ) \Bigr ] ~.
\label{mvib}
\ea
The parameters $\xi_1$ and $\xi_2$ in Eq.~(\ref{mvib}) are linear 
combinations of those in Eq.~(\ref{ms3}). In particular, since now  
$v_0=-\xi_1 R^2 \neq 0$, the corresponding eigenfunctions are collective 
in the sense that they are spread over many different oscillator shells. 

\begin{figure}
\centering
\setlength{\unitlength}{1pt}
\begin{picture}(350,200)(40,40)
\thinlines
\put ( 65, 50) {$v_1$-vibration}
\put ( 50,100) {\circle*{10}}
\put (130,100) {\circle*{10}}
\put ( 90,180) {\circle*{10}}
\put ( 50,100) {\line ( 4, 3){40}} 
\put (130,100) {\line (-4, 3){40}} 
\put ( 90,180) {\line ( 0,-1){50}} 
\thicklines
\put ( 50,100) {\vector(-4,-3){12}}
\put (130,100) {\vector( 4,-3){12}}
\put ( 90,180) {\vector( 0, 1){15}}
\thinlines
\put (195, 50) {$v_{2a}$-vibration}
\put (180,100) {\circle*{10}}
\put (260,100) {\circle*{10}}
\put (220,180) {\circle*{10}}
\put (180,100) {\line ( 4, 3){40}} 
\put (260,100) {\line (-4, 3){40}} 
\put (220,180) {\line ( 0,-1){50}} 
\thicklines
\put (180,100) {\vector( 4,-3){12}}
\put (260,100) {\vector(-4,-3){12}}
\put (220,180) {\vector( 0, 1){15}}
\thinlines
\put (325, 50) {$v_{2b}$-vibration}
\put (310,100) {\circle*{10}}
\put (390,100) {\circle*{10}}
\put (350,180) {\circle*{10}}
\put (310,100) {\line ( 4, 3){40}} 
\put (390,100) {\line (-4, 3){40}} 
\put (350,180) {\line ( 0,-1){50}} 
\thicklines
\put (310,100) {\vector(-1,-2){ 6}}
\put (390,100) {\vector(-1, 2){ 6}}
\put (350,180) {\vector( 1, 0){15}}
\end{picture}
\caption[]{\small  
Vibrations of the string-like configuration of Fig.~\ref{geometry}}
\label{vibrations}
\end{figure}
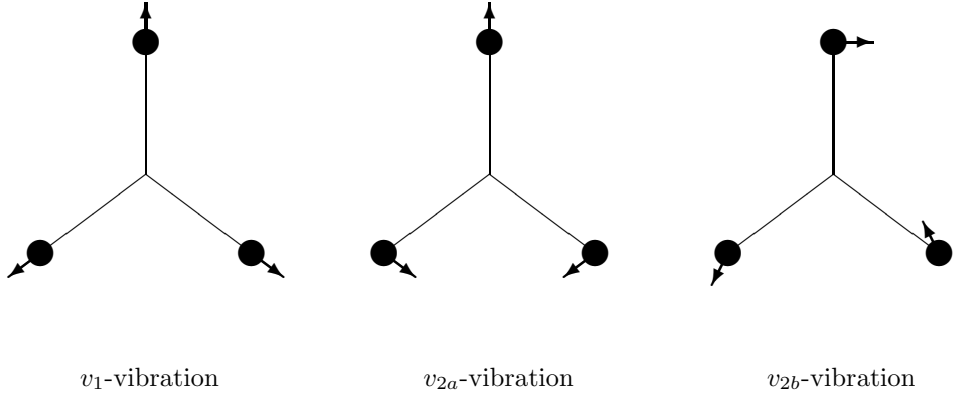

Although the mass spectrum and corresponding eigenfunctions of $\hat M^2$ 
can be obtained numerically by diagonalization, approximate solutions 
exist in the limit of a large model space ($N \rightarrow \infty$)  
which can be used to gain insight into its physical content. 
In the large $N$ limit the mass operator of Eq.~(\ref{mvib}) reduces 
to leading order in $N$ to a harmonic form, and its eigenvalues are 
given by \cite{BIL}
\ba
M^2_{\rm vib} \;=\; \kappa_1 \, v_1  + \kappa_2 \, (v_{2a}+v_{2b}) ~,
\ea
with 
\ba
\kappa_1 \;=\; 4N \xi_1 \, R^2 ~, \hspace{1cm} 
\kappa_2 \;=\; 4N \xi_2 \, R^2 /(1+R^2) ~. 
\ea
The vibrational mass operator of Eq.~(\ref{mvib}) 
has a very simple physical interpretation. Its spectrum has three 
fundamental vibrations (see Figure~\ref{vibrations}). 
The $v_1$-vibration is the
symmetric stretching vibration along the direction of the strings
(breathing mode), while the $v_{2a}$- and the $v_{2b}$-vibrations denote
bending vibrations. The latter two are 
degenerate in the case of three identical objects. 
QCD-based arguments suggest that while the string is soft towards 
stretching, it is hard towards bending and thus one expects the 
$v_2$-vibration to lie higher than the $v_1$-vibration. 
The spectrum consists of a series of vibrational excitations 
characterized by the labels ($v_1,v_2)=(v_1,v_{2a}+v_{2b})$ and a 
tower of rotational excitations built on top of each vibration. 
The rotational states for each type of vibration are  
those of an oblate symmetric top. 

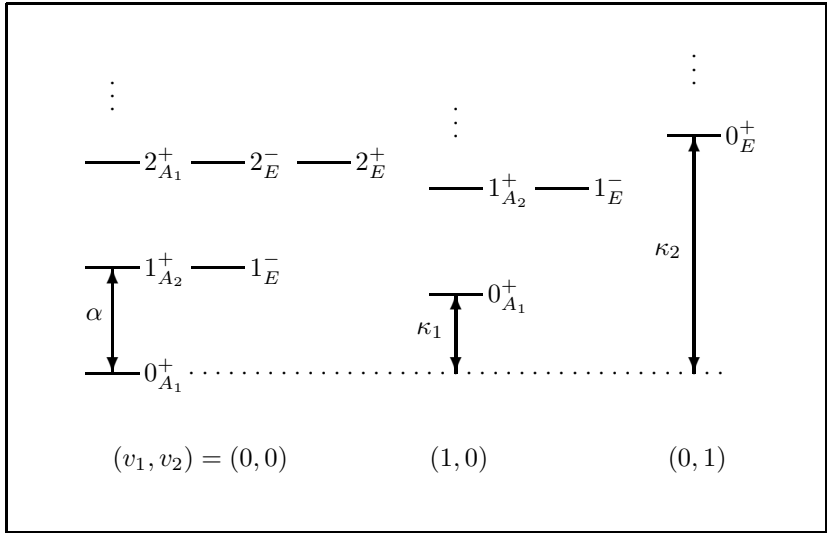
\begin{figure}
\centering
\setlength{\unitlength}{1.0pt}
\begin{picture}(310,200)(0,0)
\thinlines
\put (  0,  0) {\line(1,0){310}}
\put (  0,200) {\line(1,0){310}}
\put (  0,  0) {\line(0,1){200}}
\put (310,  0) {\line(0,1){200}}
\thicklines
\put ( 30, 60) {\line(1,0){20}}
\put ( 30,100) {\line(1,0){20}}
\put ( 30,140) {\line(1,0){20}}
\put ( 70,100) {\line(1,0){20}}
\put ( 70,140) {\line(1,0){20}}
\put (110,140) {\line(1,0){20}}
\put ( 40, 60) {\vector(0, 1){40}}
\put ( 40,100) {\vector(0,-1){40}}
\multiput ( 40,160)(0,5){3}{\circle*{0.1}}
\multiput ( 70, 60)(5,0){41}{\circle*{0.1}}
\thinlines
\put ( 30, 80) {$\alpha$}
\put ( 40, 25) {$(v_1,v_2)=(0,0)$}
\put ( 52, 57) {$0^+_{A_1}$}
\put ( 52, 97) {$1^+_{A_2}$}
\put ( 92, 97) {$1^-_E$}
\put ( 52,137) {$2^+_{A_1}$}
\put ( 92,137) {$2^-_E$}
\put (132,137) {$2^+_E$}
\thicklines
\put (160, 90) {\line(1,0){20}}
\put (160,130) {\line(1,0){20}}
\put (200,130) {\line(1,0){20}}
\put (170, 60) {\vector(0, 1){30}}
\put (170, 90) {\vector(0,-1){30}}
\multiput (170,150)(0,5){3}{\circle*{0.1}}
\thinlines
\put (155, 75) {$\kappa_1$}
\put (160, 25) {$(1,0)$}
\put (182, 87) {$0^+_{A_1}$}
\put (182,127) {$1^+_{A_2}$}
\put (222,127) {$1^-_E$}
\thicklines
\put (250,150) {\line(1,0){20}}
\put (260, 60) {\vector(0, 1){90}}
\put (260,150) {\vector(0,-1){90}}
\multiput (260,170)(0,5){3}{\circle*{0.1}}
\thinlines
\put (245,105) {$\kappa_2$}
\put (250, 25) {$(0,1)$}
\put (272,147) {$0^+_E$}
\end{picture}
\caption[]{\small
Schematic representation of the radial excitations of $q^3$ baryons  
in a stringlike collective model.  
The masses are calculated using Eq.~(\ref{eorb}) 
with $\kappa_1>0$, $\kappa_2>0$ and $\alpha>0$.} 
\label{top1}
\end{figure}

The occurrence of linear Regge trajectories suggests, that one should 
add a term linear in $L$ to the mass operator
\ba
M^2_{\rm orb} \;=\; \kappa_1 \, v_1  + \kappa_2 \, v_2    
+ \alpha \, L ~.
\label{eorb}
\ea
A schematic spectrum of the stringlike collective model is presented 
in Fig.~\ref{top1}. A comparison with the mass spectrum of Fig.~\ref{hosc1} 
shows that whereas for the harmonic oscillator the excited $L^{\pi}=0^+$ 
states belong to the two-phonon ($n=2$) multiplet, in the stringlike model
they correspond to one-phonon vibrational excitations and are the 
bandheads of these fundamental vibrations.

\subsection{Wave functions}

The full algebraic structure is obtained by combining the spatial 
part ${\cal G}_{\rm orb}$ of Eq.~(\ref{u7}) with the internal 
spin-flavor-color part ${\cal G}_{\rm sfc}$ of Eq.~(\ref{gint}) 
\ba
{\cal G} \;=\; {\cal G}_{\rm orb} \otimes {\cal G}_{\rm sfc} 
\;=\; U(7) \otimes SU_{\rm sf}(6) \otimes SU_{\rm c}(3) ~. 
\ea
The baryon wave function is obtained by combining the spin-flavor part 
with the color and orbital parts in such a way that the total wave function 
is a color-singlet, and that the three quarks satisfy the Pauli 
principle, {\it i.e.} are antisymmetric under any permutation of the three 
quarks. Since the color-singlet part of the baryon wave function is 
antisymmetric ($t=A_2$, see Table~\ref{bstates}), 
the orbital-spin-flavor part has to be symmetric ($t=A_1$)
\ba
\psi_{A_2} \;=\; \left[ \psi^{\rm c}_{A_2} \times 
\psi^{\rm osf}_{A_1} \right]_{A_2} ~, 
\label{wfbaryon}
\ea
which means that the permutation symmetry of the spatial wave function is 
the same as that of the spin-flavor part (see Table~\ref{d3baryon})
\ba
\psi^{\rm osf}_{A_1} \;=\; \left[ \psi^{\rm o}_t \times 
\psi^{\rm sf}_t \right]_{A_1} ~, 
\ea
with $t=A_1$, $E$, $A_2$. 
The square brackets $[\cdots]$ denote the tensor 
coupling under the point group $D_3$. 

\begin{table}
\centering
\caption[]{\small Discrete symmetry of $q^3$ baryon states}
\vspace{15pt}
\label{d3baryon}
\begin{tabular}{ccccc}
\hline
& & & & \\
$\psi$ & $\psi^{\rm c}$ & $\psi^{\rm osf}$ & $\psi^{\rm o}$ 
& $\psi^{\rm sf}$ \\
& & & & \\
\hline
& & & & \\
$A_2$ & $A_2$ & $A_1$ & $A_1$ & $A_1$ \\
      &       &       & $E$   & $E$   \\
      &       &       & $A_2$ & $A_2$ \\
& & & & \\
\hline
\end{tabular}
\end{table}

In the more conventional notation, the 
total baryon wave function is expressed as 
\ba
\left| \Psi \right> \;=\; \left| \, ^{2S+1}\mbox{dim}\{SU_{\rm f}(3)\}_J \, 
[\mbox{dim}\{SU_{\rm sf}(6)\},L^P] \, \right> ~,
\label{baryonwf}
\ea
where $L$, $S$ and $J$ are the orbital angular momentum, the spin and the 
total angular momentum $\vec{J}=\vec{L}+\vec{S}$. As an example, the 
wave function of the nucleon is given by 
\ba
\left| \Psi_N \right> \;=\; \left| N: \, ^{2}8_{1/2} \, [56,0^+] \, \right> ~, 
\label{nucleonwf}
\ea
and that of the $\Delta$ resonance by
\ba
\left| \Psi_{\Delta} \right> \;=\; 
\left| \Delta: \, ^{4}10_{3/2} \, [56,0^+] \, \right> ~. 
\label{deltawf}
\ea

\subsection{Mass spectrum}

The mass spectrum of the baryon resonances is 
characterized by the lowlying N(1440) resonance with $J^P=1/2^+$ 
(the socalled Roper resonance), whose mass is smaller than that 
of the first excited negative parity resonances, and the occurrence 
of linear Regge trajectories. The Roper resonance has 
the same quantum numbers as the nucleon of Eq.~(\ref{nucleonwf}), 
but is associated with the first excited $L_t^P=0^+_{A_1}$ state.
In the harmonic oscillator the first excited $L_t^P=0^+_{A_1}$ state 
belongs to the $n=2$ positive parity multiplet which lies above 
the first excited negative parity state with $n=1$ (see 
Fig.~\ref{hosc1}), whereas the data show the opposite. 
In the stringlike collective model, the Roper resonance is a 
vibrational excitation whose mass is independent of that of the 
negative parity states which are interpreted as rotational 
excitations. 

\begin{figure}
\hspace{-3cm} 
\centerline{\hbox{
\epsfig{figure=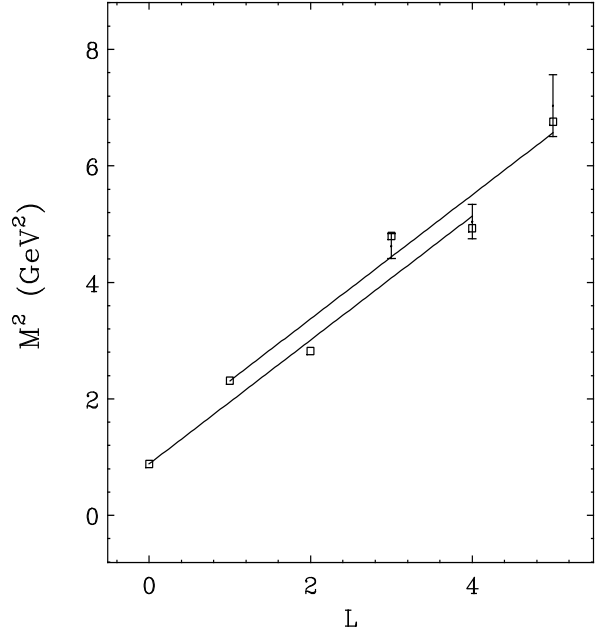,height=0.75\textwidth,width=1.0\textwidth,angle=180} }}
\caption[]{\small 
Regge trajectories for the positive parity resonances 
$| ^{2}8_{J=L+1/2} \, [56,L^+] \rangle$ with $L=0,2,4$ and 
the negative parity resonances  
$| ^{2}8_{J=L+1/2} \, [70,L^-] \rangle$ with $L=1,3,5$. 
The lines represent the result for the stringlike collective model.}
\label{Regge}
\end{figure}

Furthermore, the data show that the mass-squared of the resonances 
depends linearly on the orbital angular momentum $M^2 \propto L$. 
The resonances belonging to such a Regge trajectory have the same 
quantum numbers with the exception of $L$. The trajectories for the 
positive parity resonances $| ^{2}8_{J=L+1/2} \, [56,L^+] \rangle$ 
with $L=0,2,4$ and for the negative parity resonances  
$| ^{2}8_{J=L+1/2} \, [70,L^-] \rangle$ with $L=1,3,5$ are shown  
in Fig.~\ref{Regge}. The slope of the Regge trajectories is almost 
the same for baryons $\alpha_B=1.068$ (GeV)$^2$ \cite{BIL} and for 
mesons $\alpha_M=1.092$ (GeV)$^2$ \cite{meson}. Such a behavior is 
also expected on basis of soft QCD strings in which the strings 
elongate as they rotate \cite{soft}. 
The splitting of the rotational states in the harmonic oscillator 
is hard to reconcile with linear Regge trajectories. 

In the stringlike collective model it is straightforward 
to reproduce the relative mass of the Roper resonance and the 
occurrence of linear Regge trajectories. 
The experimental mass spectrum of baryon resonances is  
analyzed in terms of the mass formula
\ba
M^2 \;=\; M_0^2 + M^2_{\rm orb} + M^2_{\rm sf} ~,
\label{mass1}
\ea
where the orbital part is taken from Eq.~(\ref{eorb}) and  
the spin-flavor part is expressed in a G\"ursey-Radicati form 
\cite{GR}, {\it i.e.} in terms of Casimir invariants of the spin-flavor 
group $SU_{\rm sf}(6)$ and its subgroups \cite{BIL} 
\ba
M^2_{\rm sf} \;=\; a \, \left< C_{2SU_{\rm sf}(6)} \right> 
   + b \, \left< C_{2SU_{\rm f}(3)} \right> 
   + c \, S(S+1) + d \, Y + e \, Y^2 + f \, I(I+1) ~.
\label{mass2}
\ea
The explicit expressions of the eigenvalues of the Casimir operators 
of the spin-flavor and flavor groups can be found in \cite{BIL}.  
The coefficient $M^2_0$ is determined by the nucleon mass.  
The remaining nine coefficients are obtained in a simultaneous 
fit to the 48 three and four star resonances which have 
been assigned as octet and decuplet states. A good overall 
fit is found with an r.m.s. deviation of $\delta=33$ MeV. 
The values of the parameters are given in Table~\ref{par}. 

\begin{table}
\centering
\caption[]{Values of the parameters in the mass formula of 
Eqs.~(\protect\ref{mass1}) and~(\protect\ref{mass2}) in GeV$^2$}
\label{par}
\vspace{15pt}
\begin{tabular}{cr}
\hline
& \\
Parameter & Ref.~\protect\cite{BIL} \\
& \\
\hline
& \\
$\kappa_1$ &   1.204 \\
$\kappa_2$ &   1.460 \\
$\alpha$   &   1.068 \\
$a$        & --0.041 \\
$b$        &   0.017 \\
$c$        &   0.130 \\
$d$        & --0.449 \\
$e$        &   0.016 \\
$f$        &   0.042 \\
& \\
$\delta$(MeV) & 33 \\
& \\
\hline
\end{tabular}
\end{table}

The stringlike collective model provides a good overall description 
of both positive and negative baryon resonances. Table~\ref{nucdel} 
shows the results for the nucleon and $\Delta$ families. 
There is no need for an additional energy shift for the positive 
parity states and another one for the negative parity states, 
as in the relativized quark model \cite{rqm}. 
In addition to the resonances presented in the table,  
there are many more states calculated than have been observed so far, 
especially in the nucleon sector. The lowest socalled 
`missing' resonances correspond mostly to the unnatural parity states 
with $L^P=1^+$, $2^-$, which are decoupled both in electromagnetic 
and strong decays, and hence difficult to observe. 

\begin{table}
\centering
\caption[]{Mass spectrum of nonstrange baryon resonances 
in the stringlike (oblate top) model. The masses are given in MeV. 
The experimental values are taken from \protect\cite{PDG}.}
\label{nucdel} 
\vspace{15pt} 
\begin{tabular}{llcccr}
\hline
& & & & & \\
Baryon $L_{2I,2J}$ & Status & Mass 
& State & ($v_1,v_2$) & $M_{\mbox{calc}}$ \\
& & & & & \\
\hline
& & & & & \\
$N( 939)P_{11}$   & **** & 939       
& $^{2}8_{ 1/2}[56,0^+]$ & (0,0) &  939 \\
$N(1440)P_{11}$   & **** & 1430-1470 
& $^{2}8_{ 1/2}[56,0^+]$ & (1,0) & 1444 \\
$N(1520)D_{13}$   & **** & 1515-1530 
& $^{2}8_{ 3/2}[70,1^-]$ & (0,0) & 1563 \\
$N(1535)S_{11}$   & **** & 1520-1555 
& $^{2}8_{ 1/2}[70,1^-]$ & (0,0) & 1563 \\
$N(1650)S_{11}$   & **** & 1640-1680 
& $^{4}8_{ 1/2}[70,1^-]$ & (0,0) & 1683 \\
$N(1675)D_{15}$   & **** & 1670-1685 
& $^{4}8_{ 5/2}[70,1^-]$ & (0,0) & 1683 \\
$N(1680)F_{15}$   & **** & 1675-1690 
& $^{2}8_{ 5/2}[56,2^+]$ & (0,0) & 1737 \\
$N(1700)D_{13}$   &  *** & 1650-1750 
& $^{4}8_{ 3/2}[70,1^-]$ & (0,0) & 1683 \\
$N(1710)P_{11}$   &  *** & 1680-1740 
& $^{2}8_{ 1/2}[70,0^+]$ & (0,1) & 1683 \\
$N(1720)P_{13}$   & **** & 1650-1750 
& $^{2}8_{ 3/2}[56,2^+]$ & (0,0) & 1737 \\
$N(2190)G_{17}$   & **** & 2100-2200 
& $^{2}8_{ 7/2}[70,3^-]$ & (0,0) & 2140 \\
$N(2220)H_{19}$   & **** & 2180-2310 
& $^{2}8_{ 9/2}[56,4^+]$ & (0,0) & 2271 \\
$N(2250)G_{19}$   & **** & 2170-2310 
& $^{4}8_{ 9/2}[70,3^-]$ & (0,0) & 2229 \\
$N(2600)I_{1,11}$ &  *** & 2550-2750 
& $^{2}8_{11/2}[70,5^-]$ & (0,0) & 2591 \\
& & & & & \\
$\Delta(1232)P_{33}$ & **** & 1230-1234  
& $^{4}10_{3/2}[56,0^+]$ & (0,0) & 1246 \\
$\Delta(1600)P_{33}$ &  *** & 1550-1700 
& $^{4}10_{3/2}[56,0^+]$ & (1,0) & 1660 \\
$\Delta(1620)S_{31}$ & **** & 1615-1675 
& $^{2}10_{1/2}[70,1^-]$ & (0,0) & 1649 \\
$\Delta(1700)D_{33}$ & **** & 1670-1770 
& $^{2}10_{3/2}[70,1^-]$ & (0,0) & 1649 \\
$\Delta(1905)F_{35}$ & **** & 1870-1920 
& $^{4}10_{5/2}[56,2^+]$ & (0,0) & 1921 \\ 
$\Delta(1910)P_{31}$ & **** & 1870-1920 
& $^{4}10_{1/2}[56,2^+]$ & (0,0) & 1921 \\ 
$\Delta(1920)P_{33}$ &  *** & 1900-1970 
& $^{4}10_{3/2}[56,2^+]$ & (0,0) & 1921 \\ 
$\Delta(1930)D_{35}$ &  *** & 1920-1970 
& $^{2}10_{5/2}[70,2^-]$ & (0,0) & 1946 \\ 
$\Delta(1950)F_{37}$ & **** & 1940-1960 
& $^{4}10_{7/2}[56,2^+]$ & (0,0) & 1921 \\ 
$\Delta(2420)H_{3,11}$ & **** & 2300-2500 
& $^{4}10_{11/2}[56,4^+]$ & (0,0) & 2414 \\
& & & & & \\
\hline
\end{tabular}
\end{table}

\subsection{Magnetic moments}

The magnetic moment of a multiquark system is given by the 
sum of the magnetic moments of its constituent parts 
\ba
\vec{\mu} \;=\; \vec{\mu}_{\rm spin} + \vec{\mu}_{\rm orb} \;=\; 
\sum_i \mu_i (2\vec{s}_{i} + \vec{\ell}_{i}) ~, 
\ea
where $\mu_i=e_i/2m_i$, $e_i$ and $m_i$ represent the magnetic moment, 
the electric charge and the mass of the $i$-th constituent. 

The orbital-spin-flavor wave function of the ground state baryons 
is given by
\ba
\psi^{\rm osf}_{A_1} \;=\;  
\left[ \psi^{\rm o}_{A_1} \times \psi^{sf}_{A_1} \right]_{A_1} ~. 
\ea
The spin-flavor part can be expressed in terms of the 
flavor $\phi$ and spin $\chi$ wave function as 
\ba
\psi^{\rm sf}_{A_1} \;=\; \left[ \phi_{A_1} \times \chi_{A_1} \right]_{A_1} 
\;=\; \phi_{A_1} \chi_{A_1} ~,
\ea
for the decuplet baryons and 
\ba
\psi^{\rm sf}_{A_1} \;=\; \left[ \phi_{E} \times \chi_{E} \right]_{A_1} 
\;=\; \frac{1}{\sqrt{2}} \left( \phi_{E_{\rho}} \chi_{E_{\rho}} 
+ \phi_{E_{\lambda}} \chi_{E_{\lambda}} \right) ~,
\ea
for the octet baryons. 
Since the orbital wave function of the ground state baryons 
has $L^P_t=0^+_{A_1}$ (see Table~\ref{nucdel}), 
the magnetic moment only depends on the 
spin part. The magnetic moments of the $\Delta^{++}$ and the proton 
can be derived using the explicit expressions of the corresponding 
flavor and spin wave functions given in the appendices   
\ba
\mu_{\Delta^{++}} \;=\; 3\mu_u ~, \hspace{1cm} 
\mu_p \;=\; \frac{1}{3}(4\mu_u - \mu_d) ~.
\ea
Similarly, the magnetic moment of the neutron can be derived as 
\ba
\mu_n \;=\; \frac{1}{3}(4\mu_d - \mu_u) ~.
\ea
In the limit of isospin symmetry $m_u=m_d$, one recovers 
the well-known relation for the magnetic moment ratio \cite{Pais} 
\ba
\frac{\mu_n}{\mu_p} \;=\; -\frac{2}{3} ~,
\ea
which is very close to the experimental value $-0.685$.

The magnetic moments of all ground state octet and decuplet 
baryons are given in Table~\ref{mmbaryon}.  
The quark magnetic moments $\mu_u$, $\mu_d$ and $\mu_s$ are determined 
from the proton, neutron and $\Lambda$ magnetic moments to be $\mu_u=1.852$ 
$\mu_N$, $\mu_d=-0.972$ $\mu_N$ and $\mu_s=-0.613$ $\mu_N$ \cite{PDG}. 
The corresponding constituent quark masses are 
$m_u=0.338$ GeV, $m_d=0.322$ GeV, $m_s=0.510$ GeV. 
Table~\ref{mmbaryon} shows that the quark model results are in good 
agreement with the available experimental data. 

\begin{table}
\centering
\caption[]{\small 
Magnetic moments of the ground state octet and decuplet baryons in $\mu_N$. 
Experimental data are taken from \protect\cite{PDG}. 
\normalsize}
\label{mmbaryon}
\vspace{15pt}
\begin{tabular}{lcrc}
\hline
& & & \\
& $\mu_{\rm th}$ & $\mu_{\rm calc}$ & $\mu_{\rm exp}$ \\
& & & \\
\hline
& & & \\
$p$        & $(4\mu_u - \mu_d)/3$ & $ 2.793$ & $ 2.793 $          \\ 
$n$        & $(4\mu_d - \mu_u)/3$ & $-1.913$ & $-1.913 $          \\
$\Lambda$  & $\mu_s$ & $-0.613$ & $-0.613 \pm 0.004$ \\
$\Sigma^+$ & $(4\mu_u - \mu_s)/3$ & $ 2.674$ & $ 2.458 \pm 0.010$ \\
$\Sigma^0$ & $(2\mu_u + 2\mu_d - \mu_s)/3$ & $ 0.791$ &                    \\
$\Sigma^-$ & $(4\mu_d - \mu_s)/3$ & $-1.092$ & $-1.160 \pm 0.025$ \\
$\Xi^0$    & $(4\mu_s - \mu_u)/3$ & $-1.435$ & $-1.250 \pm 0.014$ \\
$\Xi^-$    & $(4\mu_s - \mu_d)/3$ & $-0.493$ & $-0.651 \pm 0.003$ \\
& & & \\
\hline
& & & \\
$\Delta^{++}$ & $3 \mu_u$ & $ 5.556$ & $5.6 \pm 1.9$ \\
$\Delta^{+}$  & $2 \mu_u + \mu_d$ & $ 2.732$ & \\
$\Delta^{0}$  & $\mu_u + 2 \mu_d$ & $-0.092$ & \\
$\Delta^{-}$  & $3 \mu_d$ & $-2.916$ & \\
$\Sigma^{\ast,+}$ & $2 \mu_u + \mu_s$ & $ 3.091$ & \\
$\Sigma^{\ast,0}$ & $\mu_u + \mu_d + \mu_s$ & $0.267$ & \\
$\Sigma^{\ast,-}$ & $2 \mu_d + \mu_s$ & $-2.557$ & \\
$\Xi^{\ast,0}$    & $\mu_u + 2 \mu_s$ & $ 0.626$ & \\
$\Xi^{\ast,-}$    & $\mu_d + 2 \mu_s$ & $-2.198$ & \\
$\Omega^{-}$  & $3 \mu_s$ & $-1.839$ & $-2.02 \pm 0.05$ \\
& & & \\
\hline
\end{tabular}
\end{table}

The magnetic moments of decuplet pentaquarks 
satisfy generalized Coleman-Glashow sum rules \cite{coleman,hong} 
\ba
\mu_{\Delta^{++}} + \mu_{\Delta^-} &=& \mu_{\Delta^+} + \mu_{\Delta^0} ~,
\nonumber\\
\mu_{\Delta^{++}} + \mu_{\Omega^-} &=& \mu_{\Sigma^{\ast,+}} + \mu_{\Xi^{\ast,0}} ~, 
\nonumber\\
\mu_{\Delta^-} + \mu_{\Omega^-} &=& \mu_{\Sigma^{\ast,-}} + \mu_{\Xi^{\ast,-}} ~, 
\ea
and
\ba
2\mu_{\Sigma^{\ast,0}} \;=\; \mu_{\Sigma^{\ast,-}} + \mu_{\Sigma^{\ast,+}}  
\;=\; \mu_{\Delta^+} + \mu_{\Xi^{\ast,-}} \;=\; \mu_{\Delta^0} + \mu_{\Xi^{\ast,0}} ~. 
\ea
The same sum rules hold for the chiral quark-soliton model 
in the chiral limit \cite{Kim1}. 

In the limit of equal quark masses $m_u=m_d=m_s=m$, the magnetic moments 
of the decuplet pentaquark states 
become proportional to their electric charges \cite{Pais} 
\ba
\mu_i \;=\; \frac{e}{2m} Q_i ~,
\label{mmb}
\ea
which means that the sum of the magnetic moments of all members of the decuplet 
vanishes identically $\sum_i \mu_i = 0$.

\section{$q^4 \bar{q}$ Pentaquarks}

The discovery of the $\Theta(1540)$ baryon with positive strangeness 
${\cal S}=+1$ by the LEPS Collaboration \cite{leps} has sparked an 
enormous amount of experimental and theoretical studies of exotic baryons. 
The NA49 Collaboration \cite{cern} reported evidence for the 
existence of another exotic baryon $\Xi^{--}(1862)$ with strangeness 
${\cal S}=-2$. The $\Theta^+$ and $\Xi^{--}$ resonances have been interpreted 
as $q^4 \bar{q}$ pentaquarks belonging to a flavor antidecuplet with quark 
structure $uudd\bar{s}$ and $ddss\bar{u}$, respectively. In addition, 
there is also evidence \cite{h1} for a heavy pentaquark $\Theta_c(3099)$ 
in which the antistrange quark in the $\Theta^+$ is replaced by an anticharm 
quark. The experimental status of the pentaquark is still unclear 
\cite{expenta}. 
Theoretical interpretations range from chiral soliton models \cite{soliton} 
which provided the motivation for the experimental searches, QCD sum rules 
\cite{sumrule}, large $N_c$ QCD \cite{largenc}, lattice QCD \cite{lattice} 
and correlated quark (or cluster) models \cite{cluster} to constituent 
quark models \cite{cqm,BGS1}. A review of the theoretical literature on  
pentaquark models can be found in \cite{thpenta}. 

In the second part of these lecture notes, I will study the properties 
of pentaquark states in a simple algebraic model, in particular the mass 
spectrum of $\Theta$ pentaquarks and the spin, parity and magnetic moment 
of the ground state pentaquark. 
As for all multiquark systems, the pentaquark wave function contains  
contributions connected to the spatial degrees of freedom  
and the internal degrees of freedom of color, flavor and spin. 
The classification of the states will be studied using symmetry 
principles which do not depend on an explicit dynamical model. 

\subsection{Internal degrees of freedom}

\begin{table}[b]
\centering
\caption[]{\small 
Allowed color spin, flavor and spin-flavor pentaquark states}
\vspace{15pt}
\label{states}
$\begin{array}{cccc}
\hline
& & & \\
& qqqq\bar{q} & \mbox{Dimension} & S_4 \sim {\cal T}_d \\
& & & \\
\hline
& & & \\
\mbox{color} & [222] & \mbox{singlet} & F_1 \\
& & & \\
\hline
& & & \\
\mbox{spin} & [5] & 6 & A_1 \\
& [41] & 4 & A_1, F_2 \\
& [32] & 2 & F_2, E   \\
& & & \\
\hline
& & & \\
\mbox{flavor} & [51] & \mbox{35-plet} & A_1 \\
& [42]  & \mbox{27-plet} & F_2 \\
& [33]  & \mbox{antidecuplet} & E \\
& [411] & \mbox{decuplet} & A_1, F_2 \\
& [321] & \mbox{octet} & F_2, E, F_1 \\
& [222] & \mbox{singlet} & F_1 \\
& & & \\
\hline
& & & \\
\mbox{spin-flavor} 
& [51111]  &  700 & A_1 \\
& [411111] &   56 & A_1, F_2 \\
& [42111]  & 1134 & F_2 \\
& [321111] &   70 & F_2, E, F_1 \\
& [33111]  &  560 & E \\
& [32211]  &  540 & F_1 \\
& [222111] &   20 & F_1, A_2 \\
& [22221]  &   70 & A_2 \\
& & & \\
\hline
\end{array}$
\end{table}

The internal degrees of freedom of the $q^4 \bar{q}$ pentaquarks are 
the same as those for the $q^3$ baryons: spin, flavor and color. 
Just as in the previous section I shall make use of the Young tableau 
technique to construct the allowed $SU_{\rm sf}(6)$ representations for 
the pentaquark $q^4\bar{q}$ system.
The pentaquark wave function should be a color singlet, and 
should be antisymmetric under any permutation of the four quarks. 
The permutation symmetry of the four-quark subsystem is characterized by 
the $S_4$ Young tableaux $[4]$, $[31]$, $[22]$, $[211]$ and $[1111]$ or, 
equivalently, by the irreducible representations of the tetrahedral group 
${\cal T}_d$ (which is isomorphic to $S_4$) as $A_1$, $F_2$, $E$, $F_1$ 
and $A_2$, respectively. For notational purposes the latter is used to 
label the discrete symmetry of the pentaquark wave functions. 
The corresponding dimensions are 1, 3, 2, 3 and 1. 
The allowed spin, flavor and spin-flavor states are obtained by 
standard group theoretic techniques (see Table~\ref{states}) 
\cite{Hamermesh,Close,Stancu} . 

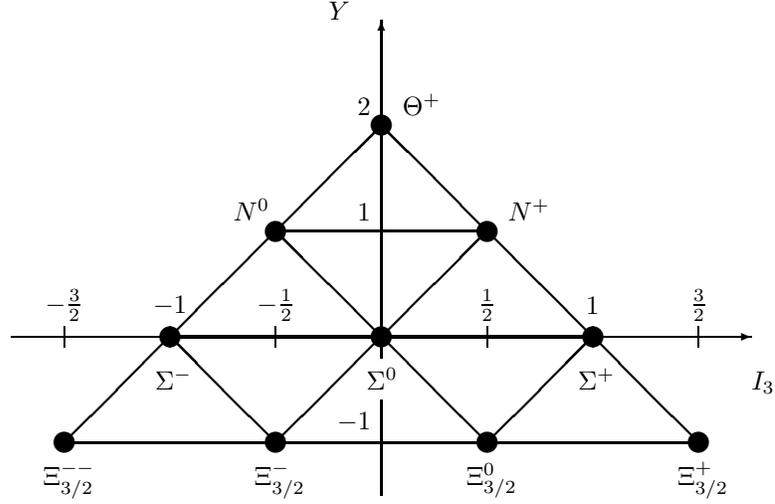
\begin{figure}
\centering
\setlength{\unitlength}{0.8pt}
\begin{picture}(350,200)(75,75)
\thinlines
\put( 75,150) {\vector(1,0){350}}
\put(250, 75) {\line(0,1){45}}
\put(250,140) {\vector(0,1){160}}
\put(100,145) {\line(0,1){10}}
\put(150,145) {\line(0,1){10}}
\put(200,145) {\line(0,1){10}}
\put(250,145) {\line(0,1){10}}
\put(300,145) {\line(0,1){10}}
\put(350,145) {\line(0,1){10}}
\put(400,145) {\line(0,1){10}}
\thicklines
\put(200,200) {\line(1,0){100}}
\put(150,150) {\line(1,0){200}}
\put(100,100) {\line(1,0){300}}

\put(100,100) {\line(1,1){150}}
\put(200,100) {\line(1,1){100}}
\put(300,100) {\line(1,1){ 50}}

\put(200,100) {\line(-1,1){ 50}}
\put(300,100) {\line(-1,1){100}}
\put(400,100) {\line(-1,1){150}}

\multiput(250,250)(100,0){1}{\circle*{10}}
\multiput(200,200)(100,0){2}{\circle*{10}}
\multiput(150,150)(100,0){3}{\circle*{10}}
\multiput(100,100)(100,0){4}{\circle*{10}}

\put(260,255){$\Theta^+$}
\put(180,205){$N^0$}
\put(310,205){$N^+$}
\put(143,125){$\Sigma^-$}
\put(243,125){$\Sigma^0$}
\put(343,125){$\Sigma^+$}
\put( 90, 80){$\Xi_{3/2}^{--}$}
\put(190, 80){$\Xi_{3/2}^{-}$}
\put(290, 80){$\Xi_{3/2}^{0}$}
\put(390, 80){$\Xi_{3/2}^+$}

\put(425,125){$I_3$}
\put(225,300){$Y$}
\put(100,165){\makebox(0,0){$-\frac{3}{2}$}}
\put(150,165){\makebox(0,0){$-1$}}
\put(200,165){\makebox(0,0){$-\frac{1}{2}$}}
\put(300,165){\makebox(0,0){$ \frac{1}{2}$}}
\put(350,165){\makebox(0,0){$ 1$}}
\put(400,165){\makebox(0,0){$ \frac{3}{2}$}}
\put(245,255){\makebox(0,0)[br]{$2$}}
\put(245,205){\makebox(0,0)[br]{$1$}}
\put(245,105){\makebox(0,0)[br]{$-1$}}
\end{picture}
\caption[]{\small Pentaquark antidecuplet}
\label{antidecuplet}
\end{figure}

In flavor space, the pentaquark states are organized into singlets, 
octets, decuplets, antidecuplets, 27-plets and 35-plets. 
It is difficult to distinguish the pentaquark flavor singlets, octets 
and decuplets from the three-quark flavor multiplets, since they have 
the same values of the hypercharge $Y$ and isospin $I$, $I_3$. The same 
observation holds for the majority of the states in the remaining flavor 
states. However, the antidecuplets, the 27-plets and 35-plets contain 
in addition exotic states with quantum numbers 
which cannot be obtained from three-quark 
configurations. These states are more easily identified experimentally 
due to the uniqueness of their quantum numbers. As an example, the exotic 
states of the antidecuplet of Fig.~\ref{antidecuplet} are:   
the $\Theta^+$ with hypercharge $Y=2$ (strangeness ${\cal S}=+1$) and 
isospin $I=I_3=0$, and the cascades $\Xi_{3/2}^+$ and $\Xi_{3/2}^{--}$ 
with hypercharge $Y=-1$ (strangeness ${\cal S}=-2$) and isospin 
$I=I_3=3/2$ and $I=-I_3=3/2$, respectively. 
The full decomposition of the spin-flavor states into spin and flavor states 
can be found in Table~6 of \cite{BGS1}. 

\subsection{Spatial degrees of freedom}

The relevant degrees of freedom for the relative motion of the 
constituent parts are provided by the Jacobi coordinates 
which are chosen as \cite{KM}
\ba
\vec{\rho} &=& \frac{1}{\sqrt{2}} 
( \vec{r}_1 - \vec{r}_2 ) ~,
\nonumber\\
\vec{\lambda} &=& \frac{1}{\sqrt{6}} 
( \vec{r}_1 + \vec{r}_2 - 2\vec{r}_3 ) ~,
\nonumber\\
\vec{\eta} &=& \frac{1}{\sqrt{12}} 
( \vec{r}_1 + \vec{r}_2 + \vec{r}_3 - 3\vec{r}_4 ) ~,
\nonumber\\
\vec{\zeta} &=& \frac{1}{\sqrt{20}} 
( \vec{r}_1 + \vec{r}_2 + \vec{r}_3 + \vec{r}_4 - 4\vec{r}_5 ) ~, 
\label{jacobi2}
\ea
where $\vec{r}_i$ ($i=1,..,4$) denote the coordinate of the $i$-th quark, 
and $\vec{r}_5$ that of the antiquark. The last Jacobi coordinate 
is symmetric under the interchange of the quark coordinates, 
and hence transforms as $A_1$ under ${\cal T}_d$ ($\sim S_4$), whereas 
the first three transform as three components of $F_2$ \cite{KM}. 

For the treatment of the spatial degrees of freedom, the same method 
is adopted as for three-quark baryons: for each 
independent relative coordinate (and its conjugate momentum) one 
introduces a dipole boson $b_i^{\dagger}$ with $L^P=1^-$ to which 
one adds an auxiliary scalar boson $s^{\dagger}$ with $L^P=0^+$  
\ba
s^{\dagger} ~, \; b^{\dagger}_{\rho,m} ~, \; b^{\dagger}_{\lambda,m} ~, \;  
b^{\dagger}_{\eta,m} ~, \; b^{\dagger}_{\zeta,m} ~, \;\;\; (m=0,\pm 1) ~.   
\label{bbp}
\ea
The scalar boson is added under the restriction that the total number of bosons 
\ba
\hat N \;=\; s^{\dagger} s + \sum_m \left( b^{\dagger}_{\rho,m} b_{\rho,m} 
+ b^{\dagger}_{\lambda,m} b_{\lambda,m} + b^{\dagger}_{\eta,m} b_{\eta,m} 
+ b^{\dagger}_{\zeta,m} b_{\zeta,m} \right) ~, 
\label{numberp}
\ea
is conserved. This procedure leads to a compact spectrum generating 
algebra for the radial (or orbital) excitations 
\ba 
{\cal G}_{\rm orb} \;=\; U(13) ~.  \label{u13}
\ea 
For a system of interacting bosons the model space is spanned by the 
symmetric irreducible representation $[N]$ of $U(13)$. 
The value of $N$ determines the size of the model space. 

The ${\cal T}_d$ permutation symmetry poses an additional constraint on the 
allowed interaction terms. The three vector bosons  
$b^{\dagger}_{\rho}$, $b^{\dagger}_{\lambda}$ and $b^{\dagger}_{\eta}$ 
transform as the three components $F_{2\rho}$, $F_{2\lambda}$ and 
$F_{2\eta}$ of the mixed symmetry representation $F_2$, while the scalar boson  
$s^{\dagger}$ and the last vector boson $b^{\dagger}_{\zeta}$ transform 
as the symmetric representation $A_1$. 
The choice of the Jacobi coordinates in Eq.~(\ref{jacobi2}) is consistent 
with the conventions used for the spin and flavor wave functions in the
appendices.  
The eigenvalues and corresponding eigenvectors can be obtained exactly by 
diagonalizing the spatial part of the mass operator in an appropriate basis. 
The radial wave functions have, by construction, good angular momentum $L$, 
parity $P$, and permutation symmetry $t=A_1$, $F_2$, $E$, $F_1$ 
or $A_2$. Moreover, the total number of bosons $N$ is conserved.  

The treatment of the orbital part depends on the choice 
of a specific dynamical model (harmonic oscillator, Skyrme, soliton, 
stringlike, hypercentral, ...). Here I consider a simple model 
in which the orbital motion of the pentaquark is limited to excitations 
up to $N=1$ quantum. The model space consists of five states: a  
ground state with $L^P=0^+$ and $A_1$ symmetry for the four quarks, 
and four excited states with $L^P=1^-$, three of which 
correspond to excitations in the relative coordinates of the four-quark 
subsystem and the fourth to an excitation in the relative coordinate 
between the four-quark subsystem and the antiquark. As a consequence of 
the permutation symmetry of the four quarks, the first three 
excitations form a degenerate triplet with three-fold $F_2$ symmetry, 
and the fourth has $A_1$ symmetry. 
In summary, the states in this simple model for the orbital motion are 
characterized by angular momentum $L$, parity $P$ and ${\cal T}_d$ 
symmetry $t$: $L^P_t=0^+_{A_1}$, $1^-_{F_2}$ and $1^-_{A_1}$. 

\subsubsection{Harmonic oscillator quark model}

The mass operator for the harmonic oscillator quark model for $q^4 \bar{q}$ 
pentaquark configurations that preserves the permutation symmetry among the 
four quarks is given by  
\ba
\hat M^2_{\rm orb} \;=\;  
\epsilon_1 \sum_m \left( b^{\dagger}_{\rho,m} b_{\rho,m} 
+ b^{\dagger}_{\lambda,m} b_{\lambda,m} 
+ b^{\dagger}_{\eta,m} b_{\eta,m} \right) 
+ \epsilon_2 \sum_m b^{\dagger}_{\zeta,m} b_{\zeta,m} ~. 
\ea
The first term comes from the three degenerate three-dimensional 
harmonic oscillators to describe the relative motion of the four 
quarks, and the second one from the three-dimensional harmonic 
oscillator for the relative motion of the antiquark with respect 
to the four-quark system. The energy eigenvalues are 
\ba
M^2_{\rm orb} \;=\; 
  \epsilon_1 \, \left( n_{1\rho}+n_{1\lambda}+n_{1\eta} \right) 
+ \epsilon_2 \, n_{2\zeta} ~, 
\ea
where $n_{1\rho}+n_{1\lambda}+n_{1\eta}=n_1$ and $n_{2\zeta}=n_2$ 
denote the number of oscillator quanta. 
The model space consists of the oscillator shells with 
$n_1+n_2=0,1,\ldots,N$. 

\begin{figure}
\centering
\setlength{\unitlength}{1.0pt}
\begin{picture}(210,200)(0,0)
\thinlines
\put (  0,  0) {\line(1,0){210}}
\put (  0,200) {\line(1,0){210}}
\put (  0,  0) {\line(0,1){200}}
\put (210,  0) {\line(0,1){200}}
\thicklines
\put ( 60, 40) {\line(1,0){20}}
\put ( 60,120) {\line(1,0){20}}
\put (140,150) {\line(1,0){20}}
\put ( 70, 40) {\vector(0, 1){80}}
\put ( 70,120) {\vector(0,-1){80}}
\put (150, 40) {\vector(0, 1){110}}
\put (150,150) {\vector(0,-1){110}}
\multiput (105,40)(5,0){15}{\circle*{0.1}}
\thinlines
\put ( 55, 80) {$\epsilon_1$}
\put (135, 95) {$\epsilon_2$}
\put ( 30, 37) {$(0,0)$}
\put ( 30,117) {$(1,0)$}
\put (110,147) {$(0,1)$}
\put ( 85, 37) {$0^+_{A_1}$}
\put ( 85,117) {$1^-_{F_2}$}
\put (165,147) {$1^-_{A_1}$}
\end{picture}
\caption[]{\small 
Schematic representation of the radial excitations 
of $q^4 \bar{q}$ pentaquarks in a harmonic oscillator model. 
The number of bosons is $N=1$. The vibrational quantum numbers  
are denoted by $(n_1,n_2)$.}
\label{hosc2}
\end{figure}
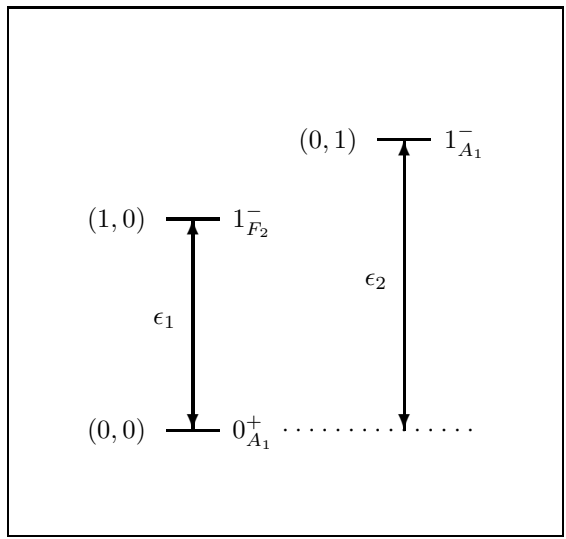

The mass spectrum for the harmonic oscillator is shown in Fig.~\ref{hosc2} 
for $N=1$ boson. The $(n_1,n_2)=(0,0)$ ground state has 
$L^P=0^+$ and $A_1$ symmetry for the four quarks. Since the orbital 
excitations are described by four relative coordinates, there are 
four excited $L^P=1^-$ states, three of which correspond to an 
excitation in the relative coordinates of the four quarks 
$(n_1,n_2)=(1,0)$, and the fourth to an excitation in the relative 
coordinate of the four-quark system and the antiquark $(0,1)$. 
As a consequence of the discrete symmetry of the four quarks, the 
first three excitations form a degenerate triplet with three-fold 
$F_2$ symmetry, and the fourth has $A_1$ symmetry.

\subsubsection{Stringlike collective model}

\begin{figure}
\hspace{-1cm} \centerline{\hbox{
\epsfig{figure=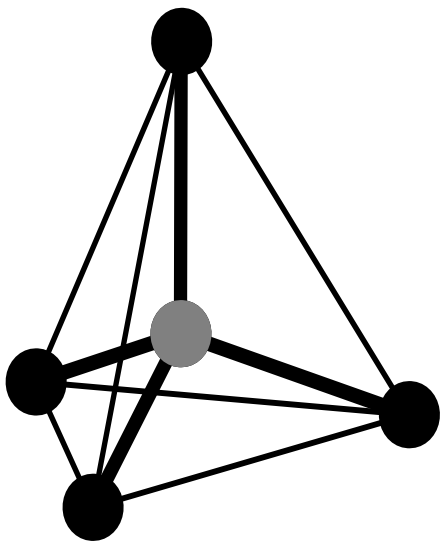,height=0.5\textwidth,width=0.5\textwidth} }}
\caption[]{\small Geometry of stringlike pentaquarks}
\label{tetrahedron}
\end{figure}

In this section I discuss a stringlike model for pentaquarks as a  
generalization of a stringlike model discussed in the previous section   
for $q^3$ baryons \cite{BIL}. In this approach, the radial excitations of 
the pentaquark are interpreted as rotations and vibrations of the 
string configuration of Fig.~\ref{tetrahedron}. 
As a consequence of the invariance of the interations under the permutation  
symmetry of the four quarks, the most favorable geometric configuration is an 
equilateral tetrahedron in which the four quarks are located at the four 
corners and the antiquark at its center \cite{stupenta}. 
This configuration was also considered in \cite{song}  
in which arguments based on the flux-tube model were used to suggest a 
nonplanar structure for the $\Theta(1540)$ pentaquark to explain its narrow 
width. In the flux-tube model, the strong color field between a pair of a 
quark and an antiquark forms a flux tube which confines them. 
For the pentaquark there would be four such flux tubes connecting the  
quarks with the antiquark. 

The vibrational spectrum of the stringlike $q^4 \bar{q}$ configuration 
with tetrahedral symmetry is characterized by four normal modes 
\ba
M^2_{\rm vib} \;=\; \kappa_1 \, v_1 
+ \kappa_2 \, \left( v_{2a}+v_{2b} \right)  
+ \kappa_3 \, \left( v_{3a}+v_{3b}+v_{3c} \right)  
+ \kappa_4 \, \left( v_{4a}+v_{4b}+v_{4c} \right) ~,
\label{mvib2}  
\ea
where $v_1$ refers to the symmetric stretching vibration (breathing mode), 
and $v_{2a}+v_{2b}=v_2$ and $v_{4a}+v_{4b}+v_{4c}=v_4$ to bending modes 
of the four-quark configuration, whereas $v_{3a}+v_{3b}+v_{3c}=v_3$ 
correspond to the relative vibration of the antiquark with respect to 
the four-quark system \cite{herzberg}. The radial excitations of the 
stringlike configuration of Fig.~\ref{tetrahedron} consists of a series 
of vibrational excitations labeled by $(v_1,v_2,v_3,v_4)$ and a tower of 
rotational excitations built on top of each vibration. 
For the simple case that is being considered here (model space with $N=1$), 
the wave functions in the stringlike model coincide with those of the 
harmonic oscillator quark model. 

\subsection{Wave functions}

The pentaquark wave function is obtained by combining the spin-flavor part 
with the color and orbital parts in such a way that the total wave function 
is a color-singlet, and that the four quarks satisfy the Pauli 
principle, {\it i.e.} are antisymmetric under any permutation of the four quarks. 
Since the color part of the pentaquark wave function is a $[222]$ singlet 
and that of the antiquark a $[11]$ anti-triplet, the color wave function of 
the four-quark configuration is a $[211]$ triplet which has $F_1$ symmetry 
under ${\cal T}_d$. The total $q^4$ wave function is antisymmetric ($A_2$), 
hence the orbital-spin-flavor part has to have $F_2$ symmetry
\ba
\psi_{A_2} \;=\; \left[ \psi^{\rm c}_{F_1} \times 
\psi^{\rm osf}_{F_2} \right]_{A_2} ~.
\label{wf}
\ea
Here the square brackets $[\cdots]$ denote the tensor coupling under the 
tetrahedral group ${\cal T}_d$. 
In Table~\ref{tdpenta}, I present the allowed spin-flavor multiplets with 
exotic pentaquarks for some lowlying orbital excitations. The exotic 
spin-flavor states associated with the state $L^P_t=0^+_{A_1}$ 
all belong to the $[42111]$ spin-flavor multiplet with $F_2$ symmetry.
The corresponding orbital-spin-flavor wave function is given by 
\ba
\psi^{\rm osf}_{F_2} \;=\;
\left[ \psi^{\rm o}_{A_1} \times \psi^{\rm sf}_{F_2} \right]_{F_2} ~.
\label{wf1} 
\ea
A radial excitation with $L^P_t=1^-_{F_2}$ gives rise to exotic 
pentaquark states of the $[51111]$, $[42111]$, $[33111]$ and 
$[32211]$ spin-flavor configurations with symmetry
$A_1$, $F_2$, $E$ and $F_1$, respectively. They are characterized by 
the orbital-spin-flavor wave functions 
\ba
\psi^{\rm osf}_{F_2} \;=\;
\left[ \psi^{\rm o}_{F_2} \times \psi^{\rm sf}_{t} \right]_{F_2} ~,
\label{wf2}
\ea
with $t=A_1$, $F_2$, $E$ and $F_1$.

\begin{table}
\centering
\caption[]{\small Discrete symmetry of exotic pentaquark states}
\vspace{15pt}
\label{tdpenta}
\begin{tabular}{cccccc}
\hline
& & & & & \\
$\psi$ & $\psi^{\rm c}$ & $\psi^{\rm osf}$ & $\psi^{\rm o}$ 
& $\psi^{\rm sf}$ & Exotic spin-flavor \\
& & & & & configuration \\
& & & & & \\
\hline
& & & & & \\
$A_2$ & $F_1$ &
 $F_2$ & $A_1$ & $F_2$ & $[42111]$ \\
& & & & & \\
$A_2$ & $F_1$ & $F_2$ & $F_2$ & $A_1$ & $[51111]$ \\
      & & & $F_2$ & $F_2$ & $[42111]$ \\
      & & & $F_2$ & $E$   & $[33111]$ \\
      & & & $F_2$ & $F_1$ & $[32211]$ \\
& & & & & \\
\hline
\end{tabular}
\end{table}

\subsection{Mass spectrum of $\Theta$ pentaquarks}

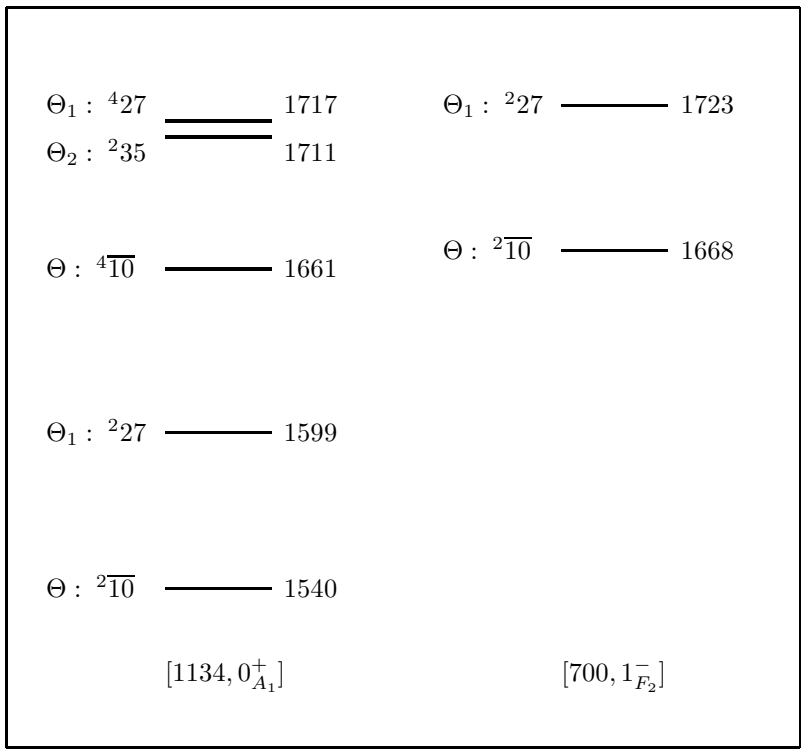
\begin{figure}
\centering
\setlength{\unitlength}{1.0pt}
\begin{picture}(300,280)(0,0)
\put(  0,  0) {\line(1,0){300}}
\put(  0,280) {\line(1,0){300}}
\put(  0,  0) {\line(0,1){280}}
\put(300,  0) {\line(0,1){280}}

\thicklines
\put( 60, 60) {\line(1,0){ 40}}
\put( 60,181) {\line(1,0){ 40}}
\put( 60,119) {\line(1,0){ 40}}
\put( 60,237) {\line(1,0){ 40}}
\put( 60,231) {\line(1,0){ 40}}
\put( 60, 25) {$[1134,0^+_{A_1}]$}

\put(210,188) {\line(1,0){ 40}}
\put(210,243) {\line(1,0){ 40}}
\put(210, 25) {$[700,1^-_{F_2}]$}

\put( 15, 57) {$\Theta: \; ^{2}\overline{10}$}
\put( 15,178) {$\Theta: \; ^{4}\overline{10}$}
\put( 15,116) {$\Theta_1: \; ^{2}27$}
\put( 15,240) {$\Theta_1: \; ^{4}27$}
\put( 15,222) {$\Theta_2: \; ^{2}35$}

\put(165,185) {$\Theta: \; ^{2}\overline{10}$}
\put(165,240) {$\Theta_1: \; ^{2}27$}

\put(105, 57) {1540}
\put(105,178) {1661}
\put(105,116) {1599}
\put(105,240) {1717}
\put(105,222) {1711}

\put(255,185) {1668}
\put(255,240) {1723}
\end{picture}
\vspace{15pt}
\caption{\small Spectrum of $\Theta$ pentaquarks. Masses are given in MeV.}
\label{theta}
\end{figure}

The classification scheme of pentaquark states discussed in the 
previous sections is based only on the fact that quarks (and antiquarks) 
have orbital, color, spin and flavor degrees of freedom. These 
states form a complete basis. The precise ordering of the pentaquark 
states in the mass spectrum depends on the choice of a specific dynamical 
model (Skyrme, CQM, Goldstone boson exchange, instanton, hypercentral, 
stringlike, ...). Since the available experimental information for 
pentaquark states is still being discussed and carefully (re)examined,  
I adopt a stringlike model in which the mass spectrum of $\Theta$ 
pentaquarks is described by the same mass formula as used for $q^3$ 
baryons in the previous section (see Eq.~(\ref{mass1})) 
\ba
M^2 \;=\; M_0^2 + M_{\rm orb}^2 + M_{\rm sf}^2 ~. 
\ea
The orbital (or radial) excitations are given by 
\ba
M_{\rm orb}^2 \;=\; M_{\rm vib}^2 + \alpha \, L ~,  
\ea
where $M_{\rm vib}^2$ describes the vibrational spectrum of a tetrahedral 
$q^4 \overline{q}$ configuration (see Eq.~(\ref{mvib2})). The rotational 
energies are given by a term linear in the orbital angular momentum $L$ 
which is responsable for the linear Regge trajectories in 
baryon and meson spectra. The spin-flavor part is expressed in the 
G\"ursey-Radicati form of Eq.~(\ref{mass2}).
The coefficients $\alpha$, $a$, $b$, $c$, $d$, $e$ and $f$ are taken 
from the previous study of baryon resonances, and the constant $M_0^2$ 
is determined by identifying the ground state exotic pentaquark with the 
recently observed $\Theta(1540)$ resonance. Since the lowest orbital states 
with $L^P_t=0^+_{A_1}$ and $1^-_{F_2}$ are interpreted as rotational states, 
for these excitations there is no contribution from the vibrational term 
$M^2_{\rm vib}$ to the mass of the corresponding pentaquark states. 
The state with $L^P_t=1^-_{A_1}$ belongs to a vibration between the four-quark 
system and the antiquark with $(v_1,v_2,v_3,v_4)=(0,0,1,0)$. 
The results for the lowest $\Theta$ pentaquarks 
(with strangeness ${\cal S}=+1$) are shown in Fig.~\ref{theta}. 

The lowest pentaquark belongs to the flavor antidecuplet 
($^{2S+1}\overline{10}$) with spin $S=1/2$ and isospin $I=0$. In the 
present calculation, the ground state pentaquark belongs to the $[42111]$ 
spin-flavor multiplet with $F_2$ symmetry, indicated in Fig.~\ref{theta} 
by its dimension 1134, and an orbital excitation $0^+$ with $A_1$ symmetry. 
Therefore, the ground state has angular momentum and parity $J^P=1/2^-$. 
The first excited state at 1599 MeV is 
an isospin triplet $\Theta_1$-state of the 27-plet ($^{2S+1}27$) 
with the same value of angular momentum and parity $J^P=1/2^-$. The lowest 
pentaquark state with positive parity occurs at 1668 MeV and belongs to the 
$[51111]$ spin-flavor multiplet (dimension 700) and an orbital 
excitation $1^-$ with $F_2$ symmetry. In the absence of a spin-orbit coupling, 
in this case there is a doublet with angular momentum and parity 
$J^P=1/2^+$, $3/2^+$. 

\subsection{Magnetic moments}

Although it may be difficult to determine the value of the magnetic moment 
experimentally, it is an essential ingredient in calculations of the photo- 
and electroproduction cross sections \cite{Nam,Zhao,photo}. 
Here I discuss the magnetic moment of negative parity antidecuplet states 
that are associated with the ground state $L^P_t=0^+_{A_1}$. They belong 
to the $[42111]$ spin-flavor multiplet with $F_2$ symmetry. 
The corresponding pentaquark wave function with angular momentum $J$ 
is given by Eqs.~(\ref{wf}) and~(\ref{wf1})
\ba
\psi^{(J)}_{A_2} &=& \left[ \psi^{\rm c}_{F_1} \times 
\left[ \psi^{\rm o}_{A_1} \times \psi^{\rm sf}_{F_2} \right]_{F_2} 
\right]^{(J)}_{A_2} 
\nonumber\\
&=& \frac{1}{\sqrt{3}} \left[ \psi^{\rm o}_{A_1} \left( 
\psi^{\rm c}_{F_{1\lambda}} \psi^{\rm sf}_{F_{2\rho}} 
- \psi^{\rm c}_{F_{1\rho}} \psi^{\rm sf}_{F_{2\lambda}} 
+ \psi^{\rm c}_{F_{1\eta}} \psi^{\rm sf}_{F_{2\eta}} \right) 
\right]^{(J)} ~. 
\label{wfneg}
\ea
The spin-flavor part can be expressed as a product of the antidecuplet 
flavor wave function $\phi_E$ and the $S=1/2$ spin wave function 
$\chi_{F_2}$ 
\ba
\psi^{\rm sf}_{F_{2\rho}} \;=\; 
\left[ \phi_{E} \times \chi_{F_2} \right]_{F_2\rho} 
&=& -\frac{1}{2} \phi_{E_{\rho}} \chi_{F_{2\lambda}} 
-\frac{1}{2} \phi_{E_{\lambda}} \chi_{F_{2\rho}} 
+\frac{1}{\sqrt{2}} \phi_{E_{\rho}} \chi_{F_{2\eta}} ~,
\nonumber\\
\psi^{\rm sf}_{F_{2\lambda}} \;=\;  
\left[ \phi_{E} \times \chi_{F_2} \right]_{F_2\lambda} 
&=& -\frac{1}{2} \phi_{E_{\rho}} \chi_{F_{2\rho}} 
+\frac{1}{2} \phi_{E_{\lambda}} \chi_{F_{2\lambda}} 
+\frac{1}{\sqrt{2}} \phi_{E_{\lambda}} \chi_{F_{2\eta}} ~,
\nonumber\\
\psi^{\rm sf}_{F_{2\eta}} \;=\;
\left[ \phi_{E} \times \chi_{F_2} \right]_{F_2\eta} 
&=& \frac{1}{\sqrt{2}} \phi_{E_{\rho}} \chi_{F_{2\rho}} 
+\frac{1}{\sqrt{2}} \phi_{E_{\lambda}} \chi_{F_{2\lambda}} ~. 
\label{wfsf}
\ea
The coefficients in Eqs.~(\ref{wfneg}) and~(\ref{wfsf}) are a consequence 
of the tensor couplings under the tetrahedral group ${\cal T}_d$ 
(Clebsch-Gordon coefficients). The total angular momentum is $J=1/2$. 

Since the orbital wave function has $L^P_t=0^+_{A_1}$, the 
magnetic moment only depends on the spin part. 
The magnetic moment of the $\Theta^+$ pentaquark can be obtained 
using the explicit form of the flavor $\phi$ and spin $\chi$ wave
functions given in the appendices \cite{BGS2}
\ba
\mu_{\Theta^+} \;=\; \frac{1}{3}(2\mu_u + 2\mu_d + \mu_s) ~.
\ea
In a similar way, the magnetic moments of the other exotic states 
of the antidecuplet, $\Xi_{3/2}^+$ and $\Xi_{3/2}^{--}$, are given by 
\cite{BGS2}
\ba
\mu_{\Xi_{3/2}^{--}} \;=\; \frac{1}{3}(\mu_u + 2\mu_d + 2\mu_s) ~,
\hspace{1cm} \mu_{\Xi_{3/2}^+} \;=\; \frac{1}{3}(2\mu_u + \mu_d + 2\mu_s) ~.
\label{mmneg}
\ea
These results are independent of the orbital wave functions, and are 
valid for any quark model in which the eigenstates have good 
$SU_{\rm sf}(6)$ spin-flavor symmetry. The values are in agreement 
with the results obtained in \cite{Liu} for the MIT bag model \cite{DS}. 

In Table~\ref{mmpenta}, I present the magnetic moments of the 
antidecuplet pentaquarks for angular momentum and parity $J^P=1/2^-$ 
together with their numerical values obtained using the same values 
of the quark magnetic moments as in the discussion of the $q^3$  
baryons. The magnetic moments of the $J^P=1/2^-$ pentaquarks are 
typically an order of magnitude smaller than the proton magnetic moment.
In Table~\ref{mmnegpar} a comparison is presented with other 
theoretical predictions for negative parity pentaquarks. 

\begin{table}
\centering
\caption[]{\small Magnetic moments of the ground state antidecuplet 
pentaquarks with $J^P=1/2^-$ in $\mu_N$}
\label{mmpenta}
\vspace{15pt}
\begin{tabular}{lcr}
\hline
& & \\
& $\mu_{\rm th}$ & $\mu_{\rm calc}$ \\
& & \\
\hline
& & \\
$\Theta^+$ & $(6\mu_u + 6\mu_d + 3\mu_s)/9$ & $ 0.382$ \\
$N^0$      & $(5\mu_u + 6\mu_d + 4\mu_s)/9$ & $ 0.108$ \\
$N^+$      & $(6\mu_u + 5\mu_d + 4\mu_s)/9$ & $ 0.422$ \\ 
$\Sigma^-$ & $(4\mu_u + 6\mu_d + 5\mu_s)/9$ & $-0.166$ \\ 
$\Sigma^0$ & $(5\mu_u + 5\mu_d + 5\mu_s)/9$ & $ 0.148$ \\ 
$\Sigma^+$ & $(6\mu_u + 4\mu_d + 5\mu_s)/9$ & $ 0.462$ \\ 
$\Xi^{--}_{3/2}$ & $(3\mu_u + 6\mu_d + 6\mu_s)/9$ & $-0.440$ \\ 
$\Xi^{-}_{3/2}$  & $(4\mu_u + 5\mu_d + 6\mu_s)/9$ & $-0.126$ \\ 
$\Xi^{0}_{3/2}$  & $(5\mu_u + 4\mu_d + 6\mu_s)/9$ & $ 0.188$ \\
$\Xi^{+}_{3/2}$  & $(6\mu_u + 3\mu_d + 6\mu_s)/9$ & $ 0.502$ \\ 
& & \\
\hline
\end{tabular}
\end{table}

\begin{table}[ht]
\centering
\caption[]{\small Comparison of magnetic moments in $\mu_N$ of exotic 
antidecuplet pentaquarks with angular momentum and parity 
$J^P=1/2^-$}
\label{mmnegpar}
\vspace{15pt}
\begin{tabular}{lccrr}
\hline
& & & & \\
Method & Ref. & $\Theta^+$ & $\Xi_{3/2}^+$ & $\Xi_{3/2}^{--}$ \\
& & & & \\
\hline
& & & & \\
Present & \protect{\cite{BGS2}} & 0.38 & 0.50 & $-0.44$ \\
& & & & \\
MIT bag & \protect{\cite{Liu}} & 0.37 & 0.45 & $-0.42$ \\
JW diquark & \protect{\cite{Nam}}    & 0.49 & & \\
$KN$ bound state & \protect{\cite{Nam}}    & 0.31 & & \\
Cluster & \protect{\cite{Zhao}}   & 0.60 & & \\
QCD sum rules & \protect{\cite{Huang}} & $0.12 \pm 0.06^{\ast}$ & & \\
              & \protect{\cite{wang1}} & $0.24 \pm 0.02^{\ast}$ & & \\
              & \protect{\cite{wang2}} & $0.18 \pm 0.01^{\ast}$ & & \\
Additive quarks & \protect{\cite{Inoue}} & $0.43$ & & $-0.41$ \\
& & & & \\
\hline
& & & & \\
\multicolumn{5}{l}{$^{\ast}$ Absolute value}
\end{tabular}
\end{table}

Finally, the magnetic moments of antidecuplet pentaquarks 
satisfy the generalized Coleman-Glashow sum rules \cite{coleman,hong} 
\ba
\mu_{\Theta^+} + \mu_{\Xi^+_{3/2}} &=& \mu_{N^+} + \mu_{\Sigma^+} ~,
\nonumber\\
\mu_{\Theta^+} + \mu_{\Xi^{--}_{3/2}} &=& \mu_{N^0} + \mu_{\Sigma^-} ~, 
\nonumber\\
\mu_{\Xi^{--}_{3/2}} + \mu_{\Xi^+_{3/2}} &=& \mu_{\Xi^-_{3/2}} + 
\mu_{\Xi^0_{3/2}} ~,
\ea
and
\ba
2\mu_{\Sigma^0} \;=\; \mu_{\Sigma^-} + \mu_{\Sigma^+} 
\;=\; \mu_{N^0} + \mu_{\Xi^0_{3/2}} 
\;=\; \mu_{N^+} + \mu_{\Xi^-_{3/2}} ~. 
\ea
The same sum rules hold for the chiral quark-soliton model 
in the chiral limit \cite{Kim2}. 

In the limit of equal quark masses $m_u=m_d=m_s=m$, the magnetic moments 
of the antidecuplet pentaquark states (denoted by $i \in \overline{10}$) 
become proportional to the electric charges 
\ba
\mu_i \;=\; \frac{1}{9} \frac{e}{2m} Q_i ~,
\label{mmq}
\ea
compared to $\mu_i = (e/2m) Q_i$ of Eq.~(\ref{mmb}) for the decuplet baryons. 
Just as for the baryon decuplet, in this limit the sum of the 
magnetic moments of all members of the antidecuplet vanishes identically 
$\sum_i \mu_i = 0$. 

\section{Summary and conclusions}

In these lecture notes, I reviewed some properties of baryons and 
pentaquarks in a stringlike collective model. The permutation symmetry 
among the quarks leads to definite geometric configurations: 
a nonlinear configuration for $q^3$ baryons in which the three 
constituent quarks are located at the corners of an equilateral 
triangle (oblate top), and a nonplanar configuration for $q^4 \bar{q}$ 
pentaquarks in which the four quarks are located at the corners of an 
equilateral tetrahedron and the antiquark at its center. In this model,  
the radial excitations of the baryons and pentaquarks are interpreted 
as rotations and vibrations of the strings.  

The algebraic structure of the model makes it possible to derive 
closed expressions for physical observables, such as masses and 
electromagnetic and strong couplings. An application to baryon 
resonances of the nucleon and delta families shows a good overall 
agreement with the available experimental data. An extension of 
the stringlike model to exotic baryons of the $\Theta$ family 
shows that the ground state pentaquark belongs to a flavor 
antidecuplet, has angular momentum and parity $J^P=1/2^-$ and,
in comparison with the proton, has a small magnetic moment. The 
width is expected to be narrow due to a large suppression in the 
spatial overlap between the pentaquark and its decay products \cite{song}.

The first report of the discovery of the pentaquark has triggered 
an enormous amount of experimental and theoretical studies of the 
properties of exotic baryons. Nevertheless, there still exist many 
doubts and questions about the existence of this state since, in 
addition to various confirmations, there is an equal amount of 
experiments in which no signal has been observed. Hence, it is of the 
utmost importance to understand the origin between these apparently 
contradictory results, and to have irrefutable proof for or against 
the existence of pentaquarks. If confirmed, the measurement of the 
quantum numbers 
of the $\Theta(1540)$ and the excited pentaquark states, especially 
the angular momentum and parity, may help to distinguish between 
different models and to gain more insight into the relevant degrees 
of freedom and the underlying dynamics that determines the properties 
of exotic baryons. If not confirmed, remains the question whether 
pentaquarks exist or not, perhaps at a different mass. Theoretically, 
there is no reason why they should not exist. However, it is difficult 
to predict the mass in a model-independent way.  

\section*{Acknowledgments}

The results presented in these lectures notes were obtained in 
collaboration with Franco Iachello, Ami Leviatan, Mauro Giannini 
and Elena Santopinto. This work is supported in part by a grant 
from CONACyT, M\'exico. 

\appendix

\section{Flavor wave functions}

The states of a $SU(3)$ flavor multiplet are labeled by the isospin $I$, 
its projection $I_3$ and the hypercharge $Y$. All other flavor states 
can be obtained by applying ladder operators in flavor space and 
using the phase convention of De Swart \cite{deSwart}. Hence for each 
flavor multiplet it is sufficient to give the wave function of one member. 

The flavor wave functions of the decuplet baryons (with $A_1$ symmetry) 
can be determined from the $\Delta^{++}$   
\ba
\phi_{A_1}(\Delta^{++}) \;=\; \left| uuu \right> ~,
\label{bar10}
\ea
and those of the octet baryons (with $E$ symmetry)  
from the proton wave function 
\ba
\phi_{E_{\rho}}(p) &=& \left. \left. \frac{1}{\sqrt{2}}
\right[ \left| udu \right> - \left| duu \right> \right] ~,
\nonumber\\
\phi_{E_{\lambda}}(p) &=& \left. \left. \frac{1}{\sqrt{6}} 
\right[ 2 \left| uud \right> - \left| udu \right> - \left| duu \right> \right] ~.
\label{bar8}
\ea

Just as for the baryons, the flavor wave functions of the antidecuplet 
pentaquarks can be obtained by applying ladder operators in flavor space 
to the $\Theta^+$ wave function (with $E$ symmetry under ${\cal T}_d$) 
\ba
\phi_{E_{\rho}}(\Theta^+) &=& - \left. \left. \frac{1}{2} \right[ 
  \left| duud \right> - \left| udud \right> 
+ \left| uddu \right> - \left| dudu \right> \right] \bar{s} ~, 
\nonumber\\
\phi_{E_{\lambda}}(\Theta^+) &=& - \left. \left. \frac{1}{2\sqrt{3}} \right[ 
  \left| duud \right> + \left| udud \right> -2 \left| uudd \right> 
+ \left| uddu \right> + \left| dudu \right> -2 \left| dduu \right> 
\right] \bar{s}~.
\label{penta}
\ea
 
\section{Spin wave functions}

The spin of $q^3$ baryons can be either $S=3/2$ or $S=1/2$ (twice) with 
$A_1$ or $E$ symmetry under $D_3$, respectively. The corresponding wave 
functions are given by 
\ba
\chi_{A_1}(q^3) \;=\; \left| \uparrow \uparrow \uparrow \right> ~,
\ea
for $S=M_S=3/2$ and 
\ba
\chi_{E_{\rho}}(q^3) &=& \left. \left. \frac{1}{\sqrt{2}} \right[ 
  \left| \uparrow \downarrow \uparrow \right> 
- \left| \downarrow \uparrow \uparrow \right> \right] ~,
\nonumber\\
\chi_{E_{\lambda}}(q^3) &=& \left. \left. \frac{1}{\sqrt{6}} \right[ 
2 \left| \uparrow \uparrow \downarrow \right> 
- \left| \uparrow \downarrow \uparrow \right> 
- \left| \downarrow \uparrow \uparrow \right> \right] ~,
\ea
for $S=M_S=1/2$. Note that the spin wave functions are related to the 
flavor wave functins of Eqs.~(\ref{bar10}) and~(\ref{bar8}) by 
interchanging $u$ by $\uparrow$ and $d$ by $\downarrow$. 
The spin states with $M_S \neq S$ are obtained 
by applying the lowering operator in spin space. 

The spin of $q^4 \bar{q}$ pentaquarks can be either $S=5/2$, $S=3/2$ 
(four times) or $S=1/2$ (five times) (see Table~\ref{states}).  
The spin wave functions of antidecuplet pentaquarks of Eqs.~(\ref{wfneg}) 
and~(\ref{wfsf}) have $S=1/2$ and $F_2$ symmetry under the tetrahedral 
group. They arise as a combination of the spin wave function for the 
four-quark system with $S=1$ and $F_2$ symmetry and that of the antiquark 
with $S=1/2$
\ba
\chi_{F_{2\alpha}}(q^4 \bar{q}) \;=\;  
 \sqrt{\frac{2}{3}} \; \chi_{F_{2\alpha}}(q^4,M_S=1) \downarrow 
-\sqrt{\frac{1}{3}} \; \chi_{F_{2\alpha}}(q^4,M_S=0) \uparrow ~, 
\ea
with $\alpha=\rho$, $\lambda$, $\eta$. The $\uparrow$ and $\downarrow$ 
represent the spin of the antiquark. 
The spin wave functions of the four-quark system are given by
\ba
\chi_{F_{2\rho}}(q^4,M_S=1) &=& \left. \left.  
-\frac{1}{\sqrt{2}} \, \right[ 
  \left| \downarrow \uparrow \uparrow \uparrow \right> 
- \left| \uparrow \downarrow \uparrow \uparrow \right> \right] ~, 
\nonumber\\
\chi_{F_{2\lambda}}(q^4,M_S=1) &=& \left. \left.  
-\frac{1}{\sqrt{6}} \, \right[
  \left| \downarrow \uparrow \uparrow \uparrow \right> 
+ \left| \uparrow \downarrow \uparrow \uparrow \right> 
-2\left| \uparrow \uparrow \downarrow \uparrow \right> \right] ~,
\nonumber\\
\chi_{F_{2\eta}}(q^4,M_S=1) &=& \left. \left.  
-\frac{1}{2\sqrt{3}} \, \right[ 
  \left| \downarrow \uparrow \uparrow \uparrow \right> 
+ \left| \uparrow \downarrow \uparrow \uparrow \right> 
+ \left| \uparrow \uparrow \downarrow \uparrow \right> 
-3\left| \uparrow \uparrow \uparrow \downarrow \right> \right] ~.
\ea
The states with projection $M_S=0$ can be obtained 
by applying the lowering operator in spin space.


\begin{thebibliography}{99}

\bibitem{stern}
I. Estermann, R. Frisch and O. Stern, 
Nature {\bf 132}, 169 (1933).

\bibitem{hofstadter}
R. Hofstadter,  
Annu. Rev. Nucl. Sci. {\bf 7}, 231 (1957).

\bibitem{dis}
J.I. Friedman and H.W. Kendall, 
Annu. Rev. Nucl. Sci. {\bf 22}, 203 (1972).

\bibitem{jones}  
M.K. Jones \textit{et al.}, 
Phys. Rev. Lett. \textbf{84}, 1398 (2000); \\
O. Gayou \textit{et al.}, 
Phys. Rev. Lett. \textbf{88}, 092301 (2002).

\bibitem{dejager}
See {\it e.g.} C.E. Hyde-Wright and K. de Jager, 
Annu. Rev. Nucl. Part. Sci. {\bf 54}, 217 (2004).

\bibitem{burkert}
See {\it e.g.} V.D. Burkert and T.-S.H. Lee,
Int. J. Mod. Phys. E {\bf 13}, 1035 (2004) [arXiv:nucl-ex/0407020]. 

\bibitem{eightfold}
M. Gell-Mann and Y. Ne'eman, 
{\it The Eightfold Way}, 
(W.A. Benjamin, New York, 1964).

\bibitem{leps} 
LEPS Collaboration, T. Nakano {\it et al.}, 
Phys. Rev. Lett. {\bf 91}, 012002 (2003). 

\bibitem{expenta}
See {\it e.g.} M. Karliner and H.J. Lipkin, 
Phys. Lett. B {\bf 597}, 309 (2004) [arXiv:hep-ph/0405002];\\
Q. Zhao and F.E. Close, J. Phys. G: Nucl. Part. Phys. {\bf 31}, L1 (2005) 
[arXiv:hep-ph/0404075];\\
K.H. Hicks, Progr. Part. Nucl. Phys., in press [arXiv:hep-ex/0504027];\\
A.R. Dzierba, C.A. Meyer and A.P. Szczepaniak, 
arXiv:hep-ex/0412077. 

\bibitem{thpenta}
See {\it e.g.} B.K. Jennings and K. Maltman, 
Phys. Rev. D {\bf 69}, 094020 (2004) [arXiv:hep-ph/0308286];\\
S.-L. Zhu, Int. J. Mod. Phys. A {\bf 19}, 3439 (2004) [arXiv:hep-ph/0406204];\\
M. Oka, Progr. Theor. Phys. {\bf 112}, 1 (2004) [arXiv:hep-ph/0406211];\\
R.L. Jaffe, Phys. Rep. {\bf 409}, 1 (2005) [arXiv:hep-ph/0409065];\\
K. Goeke, H.-C. Kim, M. Prasza{\l}owicz and G.-S. Yang, 
Progr. Part. Nucl. Phys., in press [arXiv:hep-ph/0411195].

\bibitem{dzierba}
A.R. Dzierba, D. Krop, M. Swat, S. Teige and A.P. Szczepaniak, 
Phys. Rev. D {\bf 69}, 051901 (2004) [arXiv:hep-ph/0311125];\\
K. Hicks, V. Burkert, A.E. Kudryavtsev, I.I. Strakovsky and S. Stepanyan, 
arXiv:hep-ph/0411265.

\bibitem{cern}
NA49 Collaboration, C. Alt {\it et al.}, 
Phys. Rev. Lett. {\bf 92}, 042003 (2004) [arXiv:hep-ex/0310014].

\bibitem{h1}
H1 Collaboration: A. Aktas {\it et al.}, 
Phys. Lett. B {\bf 588}, 17 (2004) [arXiv:hep-ex/0403017].

\bibitem{Hamermesh}
M. Hamermesh, 
{\it Group theory and its application to physical problems}, 
(Dover Publications, New York, 1989).

\bibitem{Close}
F.E. Close, 
{\it An introduction to quarks and partons},
(Academic Press, London, 1979)

\bibitem{Stancu}
Fl. Stancu, 
{\it Group theory in subnuclear physics}, 
(Oxford University Press, Oxford, 1996). 

\bibitem{jaffe}
R.L. Jaffe, Phys. Rev. D {\bf 15}, 267 (1977); 
{\it ibid.} {\bf 15}, 281 (1977); {\it ibid.} {\bf 17}, 1444 (1978).

\bibitem{mulders}
A.Th.M. Aerts, P.J.G. Mulders and J.J. de Swart, 
Phys. Rev. D {\bf 17}, 260 (1978); 
{\it ibid.} {\bf 21}, 1370 (1980); 
{\it ibid.} {\bf 21}, 2653 (1980). 

\bibitem{RLJ}
R.L. Jaffe, 
in {\it Proceedings of the Topical Conference on Baryon Resonances}, 
Oxford, July 5-9, 1976, Eds. R.T. Ross and D.H. Saxon. 

\bibitem{sorba}
H. H\"ogaasen and P. Sorba, 
Nucl. Phys. B {\bf 145}, 119 (1978);\\
M. de Crombrugghe, H. H\"ogaasen and P. Sorba, 
Nucl. Phys. B {\bf 156}, 347 (1979);\\
C. Roiesnel, Phys. Rev. D {\bf 20}, 1646 (1979).

\bibitem{DS}
D. Strottman, Phys. Rev. D {\bf 20}, 748 (1979). 

\bibitem{BIL}
R. Bijker, F. Iachello and A. Leviatan,
Ann. Phys. (N.Y.) {\bf 236}, 69 (1994);
{\it ibid.} {\bf 284}, 89 (2000).

\bibitem{IK}
N. Isgur and G. Karl,
Phys. Rev. D {\bf 18}, 4187 (1978);
{\it ibid.} {\bf 19}, 2653 (1979);
{\it ibid.} {\bf 20}, 1191 (1979).
 
\bibitem{meson}
F. Iachello, N.C. Mukhopadhyay and L. Zhang,
Phys. Lett. B {\bf 256}, 295 (1991); Phys. Rev. D {\bf 44}, 898 (1991).
 
\bibitem{soft}
K. Johnson and C.B. Thorn,
Phys. Rev. D {\bf 13}, 1934 (1974);\\
I. Bars and A.J. Hanson,
Phys. Rev.D {\bf 13}, 1744 (1974).

\bibitem{GR}
F. G\"ursey and L.A. Radicati ,
Phys. Rev. Lett. {\bf 13}, 173 (1964).

\bibitem{rqm}
S. Capstick and N. Isgur, 
Phys. Rev. D {\bf 34}, 2809 (1986).

\bibitem{PDG}
Particle Data Group, Phys. Lett. B {\bf 592}, 1 (2004).

\bibitem{Pais}
M.A.B. B\'eg, B.W. Lee and A. Pais, 
Phys. Rev. Lett. {\bf 13}, 514 (1964).

\bibitem{coleman}
S. Coleman and S.L. Glashow, 
Phys. Rev. Lett. {\bf 6}, 423 (1961).

\bibitem{hong}
S.-T. Hong and G.E. Brown, 
Nucl. Phys. A {\bf 580}, 408 (1994).

\bibitem{Kim1}
H.-C. Kim, M. Prasza{\l}owicz and K. Goeke, 
Phys. Rev. D {\bf 57}, 2859 (1998).

\bibitem{soliton}
D. Diakonov, V. Petrov and M. Polyakov, 
Z. Phys. A {\bf 359}, 305 (1997);\\
H. Weigel, Eur. Phys. J. A {\bf 2}, 391 (1998);\\
M. Prasza{\l}owicz, Phys. Lett. B {\bf 575}, 234 (2003) [arXiv:hep-ph/0308114];\\
J. Ellis, M. Karliner and M. Prasza{\l}owicz, 
JHEP {\bf 0405}, 002 (2004) [arXiv:hep-ph/0401127].

\bibitem{sumrule}
S.L. Zhu, Phys. Rev. Lett. {\bf 91}, 232002 (2003) [arXiv:hep-ph/0307345];\\
J. Sugiyama, T. Doi and M. Oka, Phys. Lett. B {\bf 581}, 167 (2004) 
[arXiv:hep-ph/0309271].

\bibitem{largenc}
T.D. Cohen and R.F. Lebed, Phys. Lett. B {\bf 578}, 150 (2004) [arXiv:hep-ph/0309150];\\
T.D. Cohen, Phys. Lett. B {\bf 581}, 175 (2004) [arXiv:hep-ph/0309111];\\
P.V. Pobylitsa, Phys. Rev. D {\bf 69}, 074030 (2004) [arXiv:hep-ph/0310221];\\
E. Jenkins and A.V. Manohar,
Phys. Rev. Lett. {\bf 93}, 022001 (2004) [arXiv:/0401190]; 
JHEP {\bf 0406}, 039 (2004) [arXiv:0402024];\\
D. Pirjol and C. Schat, Phys. Rev. D {\bf 71}, 036004 (2005) [arXiv:hep-ph/0408293].

\bibitem{lattice}
F. Csikor, Z. Fodor, S.D. Katz and T.G. Kov\'acs, 
JHEP {\bf 0311}, 070 (2003) [arXiv:hep-lat/0309090];\\
S. Sasaki, Phys. Rev. Lett. {\bf 93}, 152001 (2004) [arXiv:hep-lat/0310014];\\
T.-W. Chiu and T.-H. Hsieh, arXiv:hep-ph/0403020;\\
N. Mathur {\it et al.}, Phys. Rev. D {\bf 70}, 074508 (2004) [arXiv:hep-ph/0406196];\\
N. Ishii, T. Doi, H. Iida, M. Oka, F. Okiharu and H. Suganuma, 
Phys. Rev. D {\bf 71}, 034001 (2005) [arXiv:hep-lat/0408030];\\
B.G. Lasscock {\it et al.}, arXiv:hep-lat/0503008;\\
F. Csikor, Z. Fodor, S.D. Katz, T.G. Kov\'acs and B.C. Toth, 
arXiv:hep-lat/0503012;\\
C. Alexandrou and A. Tsapalis, arXiv:hep-lat/0503013;\\
T.T. Takahashi, T. Umeda, T. Onogi and T. Kunihiro, 
arXiv:hep-lat/0503019;\\
K. Holland and K.J. Juge, arXiv:hep-lat/0504007.

\bibitem{cluster}
R. Jaffe and F. Wilczek, 
Phys. Rev. Lett. {\bf 91}, 232003 (2003) [arXiv:hep-ph/0307341];\\
M. Karliner and H.J. Lipkin, 
Phys. Lett. B {\bf 575}, 249 (2003) [arXiv:hep-ph/0402260];\\ 
E. Shuryak and I. Zahed, 
Phys. Lett. B {\bf 589}, 21 (2004) [arXiv:hep-ph/0310270].

\bibitem{cqm}
Fl. Stancu, Phys. Rev. D {\bf 58}, 111501 (1998);\\
C. Helminen and D.O. Riska, 
Nucl. Phys. A {\bf 699}, 624 (2002);\\ 
A. Hosaka, Phys. Lett. B {\bf 571}, 55 (2003) [arXiv:hep-ph/0307232];\\
L.Ya. Glozman, Phys. Lett. B {\bf 575}, 18 (2003) [arXiv:hep-ph/0308232];\\
Fl. Stancu and D.O. Riska, Phys. Lett. B {\bf 575}, 242 (2003) [arXiv:hep-ph/0307010];\\
C.E. Carlson, Ch.D. Carone, H.J. Kwee and V. Nazaryan, 
Phys. Lett. B {\bf 573}, 101 (2003) [arXiv:hep-ph/0307396]; 
{\it ibid.} {\bf 579}, 52 (2004) [arXiv:hep-ph/0310038]. 

\bibitem{BGS1}
R. Bijker, M.M. Giannini and E. Santopinto, 
Eur. Phys. J. A {\bf 22}, 319 (2004) [arXiv:hep-ph/0310281].

\bibitem{KM}
P. Kramer and M. Moshinsky, 
Nucl. Phys. {\bf 82}, 241 (1966);\\
Fl. Stancu, Phys. Rev. D {\bf 58}, 111501 (1998). 

\bibitem{stupenta}
R. Bijker, M.M. Giannini and E. Santopinto, 
in {\it Nuclear Physics, Large and Small}, 
Eds. R. Bijker, R.F. Casten and A. Frank,  
AIP Conference Proceedings {\bf 726}, 181 (2004) [arXiv:hep-ph/0405195];  
Rev. Mex. F{\'{\i}}s. {\bf 50 S2}, 88 (2004) [arXiv:hep-ph/0312380]; 
arXiv:hep-ph/0409022. 
 
\bibitem{song}
X.-Ch. Song and S.-L. Zhu, Mod. Phys. Lett. A {\bf 19}, 2791 (2004) 
[arXiv:hep-ph/0403093].

\bibitem{herzberg}
G. Herzberg, {\it Molecular Spectra and Molecular Structure 
II. Infrared and Raman Spectra of Polyatomic Molecules}, 
(Krieger Publishing Company, Malabar, Florida, 1991).

\bibitem{Nam}
S.I. Nam, A. Hosaka and H.-Ch. Kim, 
Phys. Lett. B {\bf 579}, 43 (2004).

\bibitem{Zhao}
Q. Zhao, 
Phys. Rev. D {\bf 69}, 053009 (2004);
Erratum {\it ibid.} {\bf 70}, 039901 (2004) [arXiv:hep-ph/0310350];\\
Q. Zhao and J.S. Al-Khalili, 
Phys. Lett. B {\bf 582}, 91 (2004). [arXiv:hep-ph/0312348].

\bibitem{photo}
K. Nakayama and K. Tsushima, Phys. Lett. B {\bf 583}, 269 (2004) 
[arXiv:hep-ph/0311112]. 

\bibitem{BGS2}
R. Bijker, M.M. Giannini and E. Santopinto, 
Phys. Lett. B {\bf 595}, 260 (2004) [arXiv:hep-ph/0403029].

\bibitem{Liu}
Y.-R. Liu, P.-Z. Huang, W.-Z. Deng, X.-L. Chen and S.-L. Zhu, 
Phys. Rev. C {\bf 69}, 035205 (2004) [arXiv:hep-ph/0312074].

\bibitem{Huang}
P.-Z. Huang, W.-Z. Deng, X.-L. Chen and S.-L. Zhu, 
Phys. Rev. D {\bf 69}, 074004 (2004) [arXiv:hep-ph/0311108].

\bibitem{wang1}
Z.-G. Wang, W.-M. Yang and S.-L. Wan, arXiv:hep-ph/0501278.

\bibitem{wang2}
Z.-G. Wang and R.-C. Hu, arXiv:hep-ph/0504273.

\bibitem{Inoue}
T. Inoue, V.E. Lyubovitskij, Th. Gutsche and A. Faessler, 
Progr. Theor. Phys. {\bf 113}, 801 (2005) [arXiv:hep-ph/0408057].

\bibitem{Kim2}
H.-C. Kim and M. Prasza{\l}owicz, 
Phys. Lett. B {\bf 585}, 99 (2004)  
[arXiv:hep-ph/0308242].

\bibitem{deSwart}
J.J. de Swart,
Rev. Mod. Phys. {\bf 35}, 916 (1963).

\end{thebibliography}
\end{document}